\documentclass[journal]{IEEEtran}

\usepackage[T1]{fontenc}
\usepackage[utf8]{inputenc}
\usepackage{cite}
\usepackage{amsmath,amssymb,amsfonts}
\usepackage{graphicx}
\usepackage{booktabs}
\usepackage{multirow}
\usepackage{makecell}
\usepackage{tabularx}
\usepackage{array}
\usepackage{xcolor}
\usepackage{listings}
\usepackage{pifont}
\usepackage{tikz}
\usetikzlibrary{arrows.meta}
\usepackage{microtype}
\usepackage{url}
\usepackage{algorithm}
\usepackage{algpseudocode}
\usepackage[colorlinks=true,linkcolor=ftblue!70!black,citecolor=ftblue!70!black,urlcolor=ftblue!70!black]{hyperref}
\usepackage[capitalize,noabbrev]{cleveref}
\crefname{figure}{Fig.}{Figs.}
\Crefname{figure}{Fig.}{Figs.}
\crefname{section}{Section}{Sections}
\crefname{table}{Table}{Tables}
\crefname{equation}{Eq.}{Eqs.}

\graphicspath{{figures/}}

\definecolor{ftblue}{HTML}{2a78d6}
\definecolor{statgreen}{HTML}{1baf7a}
\definecolor{dynviolet}{HTML}{6250d6}
\definecolor{inkgray}{HTML}{52514e}
\definecolor{boxgray}{HTML}{f3f2ef}

\newcommand{\cmark}{\ding{51}}
\newcommand{\xmark}{\ding{55}}
\newcommand{\pmark}{$\circ$}
\newcommand{\name}{\textup{\textsc{Bridge}}}
\newcommand{\detm}{\textup{\textsc{Bridge-Det}}}
\newcommand{\matm}{\textup{\textsc{Bridge-Match}}}
\newcommand{\frz}{\textup{\textsc{Frozen}}}
\newcommand{\sd}{s$\rightarrow$d}
\newcommand{\ds}{d$\rightarrow$s}
\newcommand{\Rk}[1]{R@#1}
\newcommand{\splitR}{{\normalfont\itshape split\_random}}
\newcommand{\splitF}{{\normalfont\itshape split\_family}}
\newcommand{\codeurl}{\url{https://github.com/Mouatez/bridge-static-dynamic}}
\newcommand{\finding}[1]{\par\smallskip\noindent\textbf{Finding~#1.}~}

\begin{document}

\title{One Space, Two Views: Aligning Static and Dynamic Android Malware Analysis with Contrastive Text Embeddings for Cross-View Forensic Linking}

\author{ElMouatez Billah Karbab%
\thanks{E. B. Karbab is with the Joaan Bin Jassim Academy for Defence Studies, Al-Khor, Qatar (e-mail: mouatez@karbab.net).}}

\markboth{IEEE Transactions on Information Forensics and Security}%
{One Space, Two Views: Aligning Static and Dynamic Android Malware Analysis}

\maketitle

\begin{abstract}
Android malware analysts collect two kinds of evidence about an app: static analysis reads the APK, and dynamic analysis runs the app in a sandbox.
The two views complement each other, but each view has its own feature space.
A detector trained on APKs cannot score a sandbox report, and an analyst who receives a report without a usable hash cannot ask which known APK produced it.
We present \name{}, a bridge between static and dynamic analysis that treats the two views as two descriptions of the same app, in the same way that vision--language models treat an image and its caption.
We write the framework API list of the APK and the normalized sandbox report as text, and we fine-tune one shared encoder (Qwen3-Embedding-0.6B with LoRA) with a cross-view contrastive loss and a supervised contrastive label loss.
In this space, a detector trained on one view scores the other view, a report finds and verifies the APK that produced it, a report is attributed to a malware family across views, and the fusion of the two views stays strong when one view is weak.
On 36,276 apps with both views, a detector trained on one view scores the other view with an AUC of 0.975--0.993, and above 0.91 on malware families not seen in training.
For forensic linking, a second model finds the correct APK of a report in the top 10 of 5,442 candidates for 47\% of the reports, against 0.7\% for the encoder without fine-tuning, and verifies report--APK pairs with an AUC of 0.983 against random pairs and 0.885 against pairs of the same family.
The code and the dataset (hashes and metadata of all apps, derived inputs and trained models) are publicly available at \codeurl{}.
\end{abstract}

\begin{IEEEkeywords}
Android malware, malware forensics, static analysis, dynamic analysis, contrastive learning, text embeddings, cross-view retrieval.
\end{IEEEkeywords}

\section{Introduction}
\label{sec:intro}

\IEEEPARstart{A}{ndroid} malware analysis has two main sources of evidence.
\emph{Static analysis} reads the app package (APK) without running it and extracts, for example, the permissions, the call graph or the framework APIs that the code references~\cite{arp2014drebin,aafer2013droidapiminer,mariconti2017mamadroid,karbab2018maldozer}.
\emph{Dynamic analysis} runs the app in an instrumented sandbox and records file access, network traffic, SMS messages, data leaks and loaded code~\cite{enck2010taintdroid,lantz2014droidbox,tam2015copperdroid,lindorfer2014andrubis,karbab2019maldy}.
The two sources see different things.
Static analysis does not see code that is packed, encrypted or loaded at run time.
Dynamic analysis sees only the code paths that run during a short emulation, and evasive malware can hide its payload from the sandbox~\cite{vidas2014evading,petsas2014rage}.

Research and practice treat the two views as separate problems, or they concatenate the features of both views in one hybrid detector~\cite{lindorfer2015marvin,kim2019multimodal,karbab2023swiftr}.
Each view has its own vocabulary, so the two feature spaces are not compatible.
This has direct costs for an analyst.
A detector trained on APKs cannot score a sandbox report, and the reverse is also true.
A hybrid detector needs both views at test time, but often only one view is available: the app does not run in the emulator, sandbox time is not available, or a partner shares only the report.
Family knowledge that is labelled in one view does not move to the other view.

\paragraph*{The forensic gap}
The most direct cost is forensic.
A sandbox report can arrive without a usable APK identity: a shared report can be stripped of hashes, the APK can be repackaged or re-signed so that its hash matches no known APK, or the hash in a third-party report can be wrong.
The analyst then needs two answers: \emph{which known APK produced this report?} and \emph{does this report really come from this APK?}
Hashes cannot answer these questions, and no feature space of one view can compare a report with an APK.
To the best of our knowledge, this paper is the first to link sandbox reports and APKs in one learned space and to evaluate the matching and verification of reports against APKs.

\begin{figure*}[t]
\centering
\resizebox{\textwidth}{!}{%
\begin{tikzpicture}[
  font=\scriptsize,
  box/.style={draw=#1,thick,rounded corners=2pt,align=center,fill=white,inner sep=3pt},
  arr/.style={-{Stealth[length=1.8mm]},thick,inkgray},
  use/.style={draw=inkgray,rounded corners=2pt,fill=boxgray,align=left,inner sep=3pt,text width=4.3cm}]
  \node[box=statgreen,minimum width=1.25cm] (apk) at (0,1.35) {\textbf{APK}\\\tiny(DEX files)};
  \node[box=statgreen,text width=2.25cm] (sx) at (2.45,1.35) {external framework\\API refs, \texttt{Class.method}};
  \node[box=statgreen,text width=1.7cm] (sb) at (4.85,1.35) {rarest-first cut\\$\le$2,048 tokens};
  \node[box=dynviolet,minimum width=1.25cm] (rep) at (0,-1.35) {\textbf{Sandbox}\\\tiny report (JSON)};
  \node[box=dynviolet,text width=2.25cm] (dx) at (2.45,-1.35) {events, normalized\\(no ids, no time)};
  \node[box=dynviolet,text width=1.7cm] (db) at (4.85,-1.35) {event bag\\$\le$1,024 tokens};
  \draw[arr] (apk) -- (sx); \draw[arr] (sx) -- (sb);
  \draw[arr] (rep) -- (dx); \draw[arr] (dx) -- (db);
  \node[draw=ftblue,very thick,rounded corners=3pt,fill=ftblue!6,text width=2.0cm,align=center,minimum height=3.4cm] (enc) at (7.3,0)
    {\textbf{One shared}\\\textbf{encoder}\\[2pt]Qwen3-Emb.-0.6B\\(frozen)\\+ LoRA $r{=}16$\\[3pt]mean pool, L2\\[3pt]instruction\\selects the view};
  \draw[arr] (sb.east) -- (sb.east -| enc.west);
  \draw[arr] (db.east) -- (db.east -| enc.west);
  \node[box=statgreen] (zs) at (9.55,1.35) {$z^{s}\in\mathbb{R}^{1024}$};
  \node[box=dynviolet] (zd) at (9.55,-1.35) {$z^{d}\in\mathbb{R}^{1024}$};
  \draw[arr] (zs.west -| enc.east) -- (zs.west);
  \draw[arr] (zd.west -| enc.east) -- (zd.west);
  \node[draw=inkgray,dashed,rounded corners=2pt,fill=white,align=center,inner sep=1.5pt,font=\scriptsize,text width=2.2cm] (loss) at (9.8,0) {%
    \textbf{training loss}\\
    $\mathcal{L}_{\mathrm{NCE}}+\textstyle\sum_v\lambda_v\mathcal{L}_{\mathrm{SC}}^{v}$\\
    \textup{\textsc{Det}}: $\lambda{=}(2,2,4)$\\
    \textup{\textsc{Match}}: $\lambda{=}0$};
  \begin{scope}[shift={(13.0,0)}]
    \draw[inkgray,rounded corners=4pt,fill=white] (-1.25,-1.45) rectangle (1.25,1.45);
    \node[font=\scriptsize\bfseries,inkgray] at (0,1.22) {one space};
    \foreach \x/\y in {-0.62/0.55,-0.45/0.25,-0.7/0.05,-0.35/0.58}{\fill[statgreen] (\x,\y) circle (1.6pt);}
    \foreach \x/\y in {-0.55/0.45,-0.38/0.12,-0.62/-0.07,-0.28/0.45}{\fill[dynviolet] (\x-0.04,\y-0.04) -- ++(0.09,0) -- ++(-0.045,0.08) -- cycle;}
    \foreach \x/\y in {0.45/-0.35,0.65/-0.6,0.3/-0.75,0.7/-0.25}{\fill[statgreen] (\x,\y) circle (1.6pt);}
    \foreach \x/\y in {0.55/-0.45,0.72/-0.72,0.38/-0.62,0.6/-0.15}{\fill[dynviolet] (\x-0.04,\y-0.04) -- ++(0.09,0) -- ++(-0.045,0.08) -- cycle;}
    \draw[inkgray,dashed,rounded corners=6pt] (-0.88,-0.22) rectangle (-0.12,0.8);
    \draw[red!70!black,dashed,rounded corners=6pt] (0.15,-0.95) rectangle (0.92,0.0);
    \node[font=\scriptsize,inkgray] at (-0.5,-0.4) {benign};
    \node[font=\scriptsize,red!70!black] at (0.53,0.2) {malware};
    \node[font=\scriptsize] at (0,-1.22) {\tikz\fill[statgreen] (0,0) circle (1.6pt); static \quad \tikz\fill[dynviolet] (0,0) -- (0.09,0) -- (0.045,0.08) -- cycle; dynamic};
  \end{scope}
  \draw[arr] (zs.east) -- (11.75,0.6);
  \draw[arr] (zd.east) -- (11.75,-0.6);
  \node[use,anchor=west] (u) at (14.75,0) {%
    \textbf{Uses of the shared space}\\[2pt]
    \textbf{S1} train on one view, score the other\\
    \textbf{S2} report only: detect, find its APK,\\\hspace{1.1em}attribute its family\\
    \textbf{S3} one view weak: robust fusion\\
    \textbf{S4} forensic link: match and verify\\\hspace{1.1em}``this report comes from this APK''};
  \draw[arr] (14.25,0) -- (u.west);
\end{tikzpicture}
}
\caption{\name{}. Each view of an app is written as text and cut to a token budget. One shared encoder with a view instruction maps both texts into one space. The training loss aligns the two views of each app (InfoNCE) and groups apps by label (SupCon); its weights $\lambda$ give the two models of this paper, \detm{} (Det) and \matm{} (Match). In the space, the two views of an app are close, and apps with the same label form one region. The space supports the analyst scenarios S1--S4 of \cref{tab:scenarios}.}
\label{fig:pipeline}
\end{figure*}
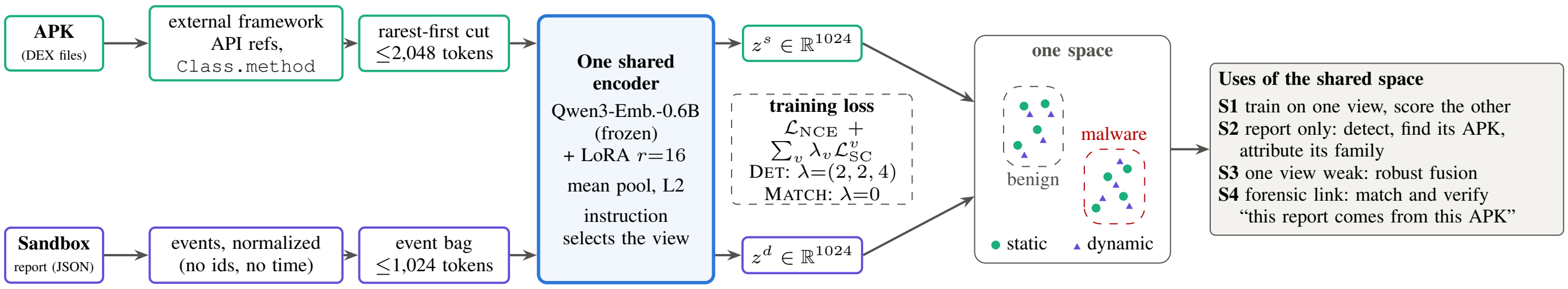

\paragraph*{Our approach}
We put both views of an app into one embedding space, in which the static fingerprint and the dynamic fingerprint of the same app are near each other (\cref{fig:pipeline}).
We call this space a \emph{bridge} between static and dynamic analysis, because in this space a classifier or a query can use either view.
We write both views as text: the static view is the list of framework APIs that the DEX code references, and the dynamic view is a normalized bag of DroidBox~\cite{lantz2014droidbox} events.
A pretrained text embedding model already knows much of this vocabulary (class names, file paths, network protocols).
We fine-tune one shared encoder, Qwen3-Embedding-0.6B~\cite{zhang2025qwen3}, with low-rank adapters (LoRA)~\cite{hu2022lora}, and a short instruction tells the encoder which view it reads.
A bidirectional InfoNCE loss~\cite{oord2018representation,radford2021learning} pulls the two views of the same app together, and supervised contrastive (SupCon) losses~\cite{khosla2020supervised} pull apps with the same label together.
We call the method \name{}.
One recipe gives two operating points: \detm{} (all loss terms) for detection and transfer, and \matm{} (InfoNCE only) for linking individual apps.

\paragraph*{What we find}
We evaluate \name{} on 36,276 apps that have both an APK and a DroidBox report: benign apps from AndroZoo~\cite{allix2016androzoo} and malware from the MalDozer collection~\cite{karbab2018maldozer} (32 families, including the Drebin~\cite{arp2014drebin} and Android Malware Genome~\cite{zhou2012dissecting} samples).
We compare against the same encoder without fine-tuning (\frz{}) and against a linear alignment of it, on a random split and on a family-disjoint split.
The frozen encoder separates the apps by view and cannot link them.
After fine-tuning, a classifier trained on one view scores the other view almost as well as a classifier trained on that view, also on unseen families.
\matm{} ranks the APK of a report at a median rank of 13 among 5,442 candidates and verifies report--APK pairs with an AUC of 0.983 against random pairs.
When one view is almost empty, the fusion stays within 0.005 AUC of the clean view (\detm{} seed 0), while the frozen fusion loses 0.06.

\paragraph*{Contributions}
\begin{itemize}
  \item \textbf{A bridge between static and dynamic analysis.} We give one contrastive text-embedding space for both views of an Android app, built with one shared encoder, view instructions and a combined InfoNCE and SupCon loss. We write both views as text, with a reduction that fits API lists, 15--70 times longer than the normalized reports at the median, into the training budget (\cref{sec:method}).
  \item \textbf{Cross-view forensic linking.} To the best of our knowledge, we are the first to introduce and evaluate the matching of sandbox reports to APKs, the verification of report--APK pairs, and cross-view family attribution, also against hard same-family negatives and with compact 64-dimensional or 128-byte binary fingerprints (\cref{sec:rq5}).
  \item \textbf{Two operating points from one loss.} We show that the SupCon terms give the class structure that detection and transfer need but remove the identity of individual apps; InfoNCE alone is about four times better at matching. The two models \detm{} and \matm{} cover the two kinds of tasks (\cref{sec:rq5,sec:ablation}).
  \item \textbf{A systematic evaluation.} We answer five research questions on 36,276 paired apps, with a family-disjoint split, three training seeds, bootstrap confidence intervals, a CCA baseline, controls for input length, time period and duplicates, and an ablation (\cref{sec:results,sec:ablation}). We share the code and the dataset at \codeurl{}: the hashes and metadata of all 36,276 apps (to download the APKs from AndroZoo), the derived inputs, the trained adapters and the fingerprints; all results are fully reproducible.
\end{itemize}

\section{Background and Threat Model}
\label{sec:background}

\paragraph*{Static view}
An APK holds one or more DEX files with the compiled code of the app, and each DEX file has a table of method references.
The references to framework classes (\texttt{android.*}, \texttt{java.*}, \texttt{javax.*}, \texttt{dalvik.*}) tell which platform functions the code can call, for example \texttt{SmsManager.sendTextMessage}.
Obfuscators do not rename framework API names, because the platform must resolve them, so API features are a standard basis for static detection~\cite{aafer2013droidapiminer,arp2014drebin,mariconti2017mamadroid,karbab2018maldozer}.
Packers, code loaded at run time from \texttt{assets/} and reflection are blind spots of this view.

\paragraph*{Dynamic view}
A sandbox installs the app in an emulator, runs it for a short time, and logs its behaviour.
We use reports of DroidBox~\cite{lantz2014droidbox}, which builds on TaintDroid~\cite{enck2010taintdroid}.
A report lists file and network operations, cryptographic operations, loaded DEX files, data leaks found by taint tracking, sent SMS messages, phone calls, services, broadcast receivers and enforced permissions.
The median report has only 17 events for benign apps and 30 events for malware, and an app can detect the emulator and stay quiet~\cite{vidas2014evading,petsas2014rage}.

\paragraph*{Complementary but incompatible}
Packing hurts the static view, and short or evasive runs hurt the dynamic view; this is the main argument for hybrid analysis~\cite{lindorfer2015marvin,kim2019multimodal}.
However, the static view names an API (\texttt{SmsManager.sendTextMessage}) and the dynamic view records its effect (\texttt{sms send 1151 msg}), so a feature-level model of one view has no input dimension for the other view.

\begin{table}[t]
\caption{Analyst scenarios and the function of the shared space.}
\label{tab:scenarios}
\centering
\footnotesize
\setlength{\tabcolsep}{3pt}
\begin{tabularx}{\linewidth}{@{}>{\raggedright}p{1.55cm} >{\raggedright}X >{\raggedright}X l@{}}
\toprule
Scenario & Available evidence & Function of the space & Sec. \\
\midrule
S1 Static-only triage & APK; no sandbox run & Detection with a classifier trained on either view & \ref{sec:rq2} \\
S2 Dynamic-only & Report; no APK (not kept, or shared by a third party) & Detection; matching to known APKs; family attribution & \ref{sec:rq2}, \ref{sec:rq5} \\
S3 Degraded view & Both views, one weak (packing, few events) & Fusion that a weak view does not harm & \ref{sec:rq3} \\
S4 Forensic link & A report and a gallery of known APKs & Matching and pair verification & \ref{sec:rq5} \\
\bottomrule
\end{tabularx}
\end{table}

\paragraph*{Threat model}
We take the view of a defender: a malware analyst, an automatic triage pipeline or a threat-intelligence team.
\Cref{tab:scenarios} lists the four scenarios that motivate a shared space.
The attacker controls the app: its code, its packaging and its run-time behaviour.
In S1--S3, the goal of the attacker is to evade detection.
In S4, the goal is to break a true link (a report does not match the APK of its app) or to forge a false one (a report matches an APK that did not produce it, for example through a poisoned report in a shared feed).
The attacker can pack or encrypt code, load code at run time, detect the emulator and add unused code.
The base encoder is public; a deployment keeps its own fine-tuned adapters, training data and labelled fingerprints private.
The training data has benign and malware labels for both views; the final models use no family labels in training.
We evaluate the natural weaknesses of each view (packed loaders, few events) and random degradation; adaptive attacks against the embedding model~\cite{grosse2017adversarial,pierazzi2020intriguing} are a direction for future work.

\section{Related Work}
\label{sec:related}

\paragraph*{Static and dynamic Android analysis}
Early static detectors use permissions and API calls~\cite{felt2011permissions,au2012pscout,backes2016demystifying,peng2012probabilistic}.
DroidAPIMiner~\cite{aafer2013droidapiminer} selects frequent malware APIs, Drebin~\cite{arp2014drebin} uses a large sparse vector with a linear classifier, and MaMaDroid~\cite{mariconti2017mamadroid,onwuzurike2019mamadroid} models abstracted API sequences with Markov chains.
Deep models followed~\cite{yuan2014droidsec,yuan2016droiddetector,xu2018deeprefiner}.
MalDozer~\cite{karbab2018maldozer,karbab2017api} treats API call sequences as text, PetaDroid~\cite{karbab2021petadroid} adds ensembles and adaptation over time, and related work studies portable and resilient fingerprinting~\cite{karbab2021book,karbab2021portable,karbab2021resilient,karbab2021resilient_arxiv,karbab2020resilientscalable} and packaging fingerprints~\cite{karbab2016dna}.
On the dynamic side, TaintDroid~\cite{enck2010taintdroid} and DroidBox~\cite{lantz2014droidbox} log data flows and system activity, DroidScope~\cite{yan2012droidscope} and CopperDroid~\cite{tam2015copperdroid} reconstruct behaviour from outside the emulator, Andrubis~\cite{lindorfer2014andrubis} analysed a million apps, and Andromaly~\cite{shabtai2012andromaly} uses on-device behaviour.
DySign~\cite{karbab2016dysign} builds dynamic fingerprints, and MalDy~\cite{karbab2019maldy,karbab2018maldy_arxiv} treats behavioural reports as text.
\name{} uses the same kinds of evidence, but its goal is not a better detector for one view: it makes static and dynamic evidence comparable.

\paragraph*{Hybrid and cross-view analysis}
Marvin~\cite{lindorfer2015marvin}, Kim et al.~\cite{kim2019multimodal}, SwiftR~\cite{karbab2023swiftr} and TriDroid~\cite{amira2021tridroid} \emph{fuse} static and dynamic features; they need all views at test time and do not make one view usable in place of the other.
Learning under privileged information~\cite{celik2018detection} and the distillation of Ngo et al.~\cite{ngo2023fast} use dynamic features only at training time to improve a static detector, which transfers knowledge in one direction only.
Multimodal pre-training aligns the views directly: MPDroid~\cite{zhang2025mpdroid} aligns graph encoders of static and dynamic call graphs, and MalCAFF~\cite{guo2026malcaff} uses contrastive pre-training between static assembly and dynamic API features of Windows executables before a fused classifier.
Li et al.~\cite{li2025multimodal} align two static modalities of Android apps, and McNdroid~\cite{kamol2026mcndroid} is a longitudinal multimodal benchmark.
\Cref{tab:related} compares these works with \name{}.
\name{} differs in three ways: it uses one pretrained \emph{text} encoder for both views; it evaluates each view \emph{alone} in the shared space (transfer in both directions and robustness to a weak view); and it evaluates the \emph{forensic} use of the space, the matching and verification of reports against APKs.
Cross-modal retrieval exists for code, between natural language and code~\cite{feng2020codebert} or between binaries of different architectures~\cite{xu2017neural,wang2022jtrans}, but not between sandbox reports and APKs.

\begin{table}[t]
\caption{Comparison with the closest work, by what each work evaluates. S/D: static/dynamic view. Space: shared space; 1-view: one view at test time; Transf.: cross-view transfer; Link: report--sample link. \cmark{} evaluated; \pmark{} one direction; \xmark{} not possible in the design; n/e possible, not evaluated; -- not applicable.}
\label{tab:related}
\centering
\footnotesize
\setlength{\tabcolsep}{2.6pt}
\begin{tabular}{@{}l l l c c c c@{}}
\toprule
Work & Platf. & Views & Space & 1-view & Transf. & Link \\
\midrule
Drebin~\cite{arp2014drebin}, MaMaDroid~\cite{mariconti2017mamadroid} & Andr. & S & -- & -- & \xmark & \xmark \\
MalDy~\cite{karbab2019maldy}, DySign~\cite{karbab2016dysign} & Andr. & D & -- & -- & \xmark & \xmark \\
Marvin~\cite{lindorfer2015marvin}, Kim~\cite{kim2019multimodal} & Andr. & S+D & \xmark & \xmark & \xmark & \xmark \\
SwiftR~\cite{karbab2023swiftr} & Multi & S+D & \xmark & \xmark & \xmark & \xmark \\
Privileged inf.~\cite{celik2018detection} & Andr. & S+D & \xmark & \pmark & \xmark & \xmark \\
Ngo et al.~\cite{ngo2023fast} & Win. & S+D & \xmark & \pmark & \xmark & \xmark \\
MalCAFF~\cite{guo2026malcaff} & Win. & S+D & \cmark & \xmark & n/e & n/e \\
MPDroid~\cite{zhang2025mpdroid} & Andr. & S+D & \cmark & \cmark & n/e & n/e \\
\textbf{\name{} (ours)} & Andr. & S+D & \cmark & \cmark & \cmark & \cmark \\
\bottomrule
\end{tabular}
\end{table}

\paragraph*{Contrastive and embedding models}
InfoNCE~\cite{oord2018representation} and SimCLR~\cite{chen2020simple} use the other samples of a batch as negatives, and CLIP~\cite{radford2021learning} applies a symmetric InfoNCE loss to image--text pairs to build a shared space for two modalities.
SupCon~\cite{khosla2020supervised} uses class labels to define the positives, and Chen et al.~\cite{chen2023continuous} use a hierarchical contrastive encoder with family labels against drift.
Multimodal models often degrade when a modality is missing at test time~\cite{ma2021smil,ma2022multimodal}.
\name{} is a CLIP-style model in which the two ``modalities'' are two views of one program.
Contrastively trained transformers~\cite{vaswani2017attention} give strong text embeddings~\cite{reimers2019sentence,wang2022text,zhang2025qwen3}; Matryoshka training~\cite{kusupati2022matryoshka} lets them be truncated, and LoRA~\cite{hu2022lora} adapts them cheaply.
Binary-similarity models learn function embeddings for search~\cite{xu2017neural,ding2019asm2vec,massarelli2019safe,li2021palmtree,wang2022jtrans,alrabaee2019bineye}, and LLMs now analyse malware directly~\cite{qian2025lamd,karbab2026asmrag,karbab2026quantifiable}.
\name{} uses an LLM-based \emph{embedding} model, which gives a fixed-size fingerprint for fast nearest-neighbour search and simple classifiers.

\paragraph*{Fingerprinting, attribution and evaluation pitfalls}
Cypider~\cite{karbab2016cypider,karbab2020scalable,karbab2020clustering} and graph-based fingerprints~\cite{karbab2025graph,amira2023survey} cluster malware without labels, and ToGather~\cite{karbab2018togather,karbab2018automatic} links samples through their network infrastructure.
These works compare samples \emph{within} one view; \name{} adds fingerprints that link a report to an APK \emph{across} views.
TESSERACT~\cite{pendlebury2019tesseract} and Arp et al.~\cite{arp2022dos} show that temporal and sampling bias inflate results, and CADE~\cite{yang2021cade} detects drifting samples.
Reference datasets~\cite{zhou2012dissecting,wei2017deep,allix2016androzoo} have their own collection periods.
We therefore use a family-disjoint split and test for length, time and duplicate shortcuts (\cref{sec:rq4}).

\section{Method}
\label{sec:method}

An \emph{app} is the unit of analysis; its \emph{APK} is the static artifact and its \emph{report} is the DroidBox output.
A \emph{fingerprint} is the normalized embedding of one view of one app, and a \emph{probe} is a logistic regression (LR) trained on fixed fingerprints.
\name{} has three steps (\cref{fig:pipeline}).
First, we write each view of an app as text (\cref{sec:inputs}).
Second, one shared encoder maps both texts to unit vectors in $\mathbb{R}^{1024}$ (\cref{sec:encoder}).
Third, a contrastive loss trains the encoder so that the two views of an app are close and the classes are separate (\cref{sec:loss}).
After training, simple downstream models use the fingerprints (\cref{sec:use}).
Supplementary Algorithms~\ref{alg:inputs}--\ref{alg:use} give the pseudocode of the three steps.

\begin{table}[t]
\caption{Data set, splits and input sizes. Token values: median / 95th percentile of the Qwen3 tokenizer; raw JSON from a sample of 1,500 reports per class. In \splitF{}, the test and validation families do not occur in training.}
\label{tab:data}
\centering
\footnotesize
\setlength{\tabcolsep}{2.5pt}
\begin{tabular}{@{}l r r r r@{}}
\toprule
 & Train & Val & Test & Total \\
\midrule
\multicolumn{5}{@{}l}{\emph{\splitR{} (70/15/15 by app, stratified)}} \\
Apps & 25,393 & 5,441 & 5,442 & 36,276 \\
\multicolumn{5}{@{}l}{\emph{\splitF{} (malware families disjoint)}} \\
Malware families & 15 & 10 & 7 & 32 \\
Apps & 24,690 & 5,707 & 5,879 & 36,276 \\
\midrule
Input size (tokens) & \multicolumn{2}{r}{Benign} & \multicolumn{2}{r}{Malware} \\
\midrule
Raw DroidBox JSON & \multicolumn{2}{r}{4,804 / 31,932} & \multicolumn{2}{r}{7,985 / 97,764} \\
Normalized event bag (\name{}) & \multicolumn{2}{r}{73 / 375} & \multicolumn{2}{r}{160 / 720} \\
Events per report & \multicolumn{2}{r}{17 / 113} & \multicolumn{2}{r}{30 / 403} \\
External framework APIs (count) & \multicolumn{2}{r}{1,106 / 2,773} & \multicolumn{2}{r}{597 / 2,129} \\
\texttt{Class.method} list, before cut & \multicolumn{2}{r}{5,000 / 12,986} & \multicolumn{2}{r}{2,457 / 9,746} \\
After the cut (\name{} input) & \multicolumn{2}{r}{2,014 / 2,032} & \multicolumn{2}{r}{1,989 / 2,026} \\
Apps changed by the cut & \multicolumn{2}{r}{75.7\%} & \multicolumn{2}{r}{53.8\%} \\
\bottomrule
\end{tabular}
\end{table}

\subsection{Text Forms of the Two Views}
\label{sec:inputs}

The raw inputs do not fit the context of a text encoder.
A DroidBox report in JSON has a median of 4,804 tokens (benign) and 7,985 tokens (malware), mostly because of hex payloads and timestamps, and the full list of framework methods has a median of 5,000 tokens for benign apps (\cref{tab:data}).
Contrastive training needs many pairs per step, so we set a budget of 2,048 tokens for the static view and 1,024 tokens for the dynamic view.

\paragraph*{Static text}
We read the \texttt{method\_ids} table of every \texttt{classes*.dex} file.
We keep a method reference only if its class has a core framework prefix and is \emph{not} defined in the APK; this removes library code that the app bundles under framework-like names.
We write each reference as \texttt{Class.method}, remove duplicates, and sort the list.
After this filter, 65\% of the apps still have more than 2,048 tokens.
For these apps, we keep the \emph{rarest} APIs first: common APIs such as \texttt{String.equals} carry little information, and rare APIs such as \texttt{SmsManager.sendTextMessage} are specific.
The rarity is the document frequency on the training apps only.
We add APIs in order of increasing document frequency until the budget is full, and then sort the kept list.

\paragraph*{Dynamic text}
We convert each DroidBox event into one line \texttt{<category> <operation> <target>}.
We keep the information that describes behaviour: the file path and operation, the network host and port, the first 48 bytes of a sent payload as printable text (HTTP method, URL path and parameters), the crypto algorithm, the loaded DEX path, the leak type and sink, the SMS number, the service, the receiver action and the permission.
We remove or replace the values that identify one app or one run, because they let the model match a report to an APK by an identifier instead of by behaviour: hashes, the APK name, timestamps, file descriptors and process ids, the package name (\texttt{<pkg>}), long hex strings and UUIDs (\texttt{<hex>}, \texttt{<uuid>}) and long digit runs (\texttt{<n>}).
We then build an order-free event \emph{bag}: the unique lines, sorted, each with its count (\texttt{x$N$}).
Only 1.0\% of the reports are longer than the dynamic budget.
\Cref{fig:example} shows both views of one malware app: the two texts describe the same behaviour, but they share almost no tokens.

\begin{figure}[t]
\centering
\footnotesize
\textbf{\textcolor{statgreen}{Static view}} (excerpt: 14 of the 181 APIs)\\[1pt]
{\fontsize{6}{7}\selectfont\ttfamily\setlength{\tabcolsep}{4pt}
\begin{tabular}{@{}l l@{}}
AlarmManager.set & SmsManager.getDefault \\
BroadcastReceiver.<init> & SmsManager.sendTextMessage \\
Context.startService & SmsMessage.createFromPdu \\
HttpURLConnection.getOutputStream & SmsMessage.getOriginatingAddress \\
HttpURLConnection.setRequestMethod & TelephonyManager.getDeviceId \\
Service.onStart & TelephonyManager.getLine1Number \\
URL.openConnection & TelephonyManager.getSubscriberId \\
\end{tabular}}\\[3pt]
\textbf{\textcolor{dynviolet}{Dynamic view}} (complete event bag: 13 unique lines)\\[-1pt]
\begin{lstlisting}
dexload /data/app/<pkg>.apk
file read /proc/<pid>/cmdline x2
file read /proc/meminfo
file write /data/data/<pkg>/shared_prefs/com.android.gallery3d_preferences.xml
file write /data/data/<pkg>/shared_prefs/com.android.providers.contacts_preferences.xml
file write /data/data/<pkg>/shared_prefs/settings.xml
leak TAINT_PHONE_NUMBER,TAINT_IMEI,TAINT_IMSI Network 23.253.126.58:80
net recv 23.253.126.58:80 x4
net send 23.253.126.58:80 POST /midlets/ins/?gensmsid=<n>&imei=3
opennet 23.253.126.58:80
receiver .MainReceiver android.intent.action.BOOT_COMPLETED
service start <pkg>.MainService
sms send 1151 msg x3
\end{lstlisting}
\caption{The two text views of one malware app (family FakeApp, training data). The static view names the APIs that read the device identifiers and send SMS messages; the dynamic view shows their effect: the IMEI, IMSI and phone number leak to a network host, and three SMS messages go to a premium number. The two texts share almost no tokens; the encoder must learn this correspondence.}
\label{fig:example}
\end{figure}

\subsection{Shared Encoder}
\label{sec:encoder}

We use Qwen3-Embedding-0.6B~\cite{zhang2025qwen3}, a 28-layer transformer with a hidden size of 1,024 that is pretrained for instruction-based text embedding.
One encoder reads both views, and an instruction prefix tells it the view (``Represent the API usage of an Android app.'' or ``Represent the runtime behavior of an Android app.'').
A shared encoder gives a token such as \texttt{Sms} one meaning in both views and does not double the model size.
The fingerprint is the mean of the last hidden states over all tokens, followed by L2 normalization:
\begin{equation}
  z^{v}_i = \bar{h}(x^{v}_i)\,/\,\lVert \bar{h}(x^{v}_i) \rVert_2 \in \mathbb{R}^{1024}, \qquad v \in \{s, d\}.
\end{equation}
We freeze the base weights and train LoRA adapters~\cite{hu2022lora} of rank 16 ($\alpha=32$, dropout 0.05) on all attention and MLP projections of every layer: about 10.1 million parameters, 1.7\% of the base model.

\subsection{Training Objective}
\label{sec:loss}

A batch has $B$ apps with static fingerprints $z^{s}_i$, dynamic fingerprints $z^{d}_i$ and binary labels $y_i$ (benign or malware).
All similarities are cosine similarities divided by a temperature $\tau=0.05$.

\paragraph*{Cross-view alignment (InfoNCE)}
The two views of the same app form a positive pair, and the other $B-1$ apps of the batch give the negatives.
As in CLIP~\cite{radford2021learning}, we use the loss in both directions:
\begin{equation}
\mathcal{L}_{\mathrm{NCE}} = -\frac{1}{2B}\sum_{i=1}^{B}\Big[
  \log\frac{e^{z^{s}_i\cdot z^{d}_i/\tau}}{\sum_{j}e^{z^{s}_i\cdot z^{d}_j/\tau}}
+ \log\frac{e^{z^{d}_i\cdot z^{s}_i/\tau}}{\sum_{j}e^{z^{d}_i\cdot z^{s}_j/\tau}}\Big].
\label{eq:nce}
\end{equation}
This term asks the model to identify \emph{the} app: the static fingerprint must be closer to its own report than to the reports of all other apps in the batch.

\paragraph*{Label structure (SupCon)}
For fingerprints $Z=\{z_1,\dots,z_N\}$ with labels $y$, the supervised contrastive loss~\cite{khosla2020supervised} is
\begin{equation}
\mathcal{L}_{\mathrm{SC}}(Z,y) = -\frac{1}{|I|}\sum_{a\in I}\frac{1}{|P(a)|}\sum_{p\in P(a)}
  \log\frac{e^{z_a\cdot z_p/\tau}}{\sum_{k\neq a}e^{z_a\cdot z_k/\tau}},
\label{eq:supcon}
\end{equation}
where $P(a)=\{p\neq a: y_p=y_a\}$ and $I$ is the set of anchors with at least one positive.
The total loss applies it across the views and to each view alone:
\begin{equation}
\begin{split}
\mathcal{L} = \mathcal{L}_{\mathrm{NCE}}
  &+ \lambda_{x}\mathcal{L}_{\mathrm{SC}}\big([z^{s};z^{d}],[y;y]\big)\\
  &+ \lambda_{d}\mathcal{L}_{\mathrm{SC}}(z^{d},y)
  + \lambda_{s}\mathcal{L}_{\mathrm{SC}}(z^{s},y).
\end{split}
\label{eq:total}
\end{equation}
The cross-view term pulls a static malware fingerprint towards the dynamic fingerprints of \emph{other} malware apps, which gives the class structure that cross-view transfer needs; the single-view terms make each view separable on its own.
We use binary labels, so training needs no family information.

\paragraph*{Two models from one recipe}
InfoNCE keeps each app distinct from all other apps.
SupCon makes all apps of a class similar, and therefore also removes part of the difference between the apps of one class.
We train two models that differ only in the weights:
\detm{} uses $(\lambda_x,\lambda_d,\lambda_s)=(2,2,4)$, selected by the validation detection AUC, and is the model for detection, transfer and fusion;
\matm{} uses InfoNCE only, $(0,0,0)$, and is the model for matching and verification.
The two models are two operating points of one loss, not the result of a search over the trade-off.

\paragraph*{Training}
Each step encodes the static and dynamic texts of a batch with the same encoder, and the fingerprints of all GPUs are gathered before the loss, so that every app sees the negatives of the whole global batch of 32 apps.
We use AdamW (learning rate $2\times10^{-4}$, weight decay 0.01, 5\% warm-up, cosine decay, clipping at 1.0) for 3 epochs.
We keep the checkpoint with the best mean validation AUC of three probes (static, dynamic, fusion) on 1,000 training apps.
\Cref{tab:hparams} lists all settings, and Supplementary Table~\ref{tab:search} gives the search.

\begin{table}[t]
\caption{Training settings. The two models differ only in the loss weights.}
\label{tab:hparams}
\centering
\footnotesize
\setlength{\tabcolsep}{3pt}
\begin{tabularx}{\linewidth}{@{}l >{\raggedright\arraybackslash}X@{}}
\toprule
Setting & Value \\
\midrule
Base model & Qwen3-Embedding-0.6B (28 layers, hidden size 1,024), frozen \\
Adapters & LoRA, rank 16, $\alpha=32$, dropout 0.05, on \texttt{q,k,v,o,gate,up,down}; 10.1\,M parameters (1.7\%) \\
Pooling & mean over all tokens, then L2 normalization \\
Max.\ length & static 2,048 tokens, dynamic 1,024 tokens \\
Loss weights & $(\lambda_x,\lambda_d,\lambda_s)$: \detm{} $(2,2,4)$, \matm{} $(0,0,0)$; SupCon on binary labels; $\tau=0.05$ \\
Optimizer & AdamW, lr $2\times10^{-4}$, weight decay 0.01, 5\% warm-up, cosine decay, clipping 1.0 \\
Batch & 4 apps per GPU $\times$ 8 GPUs = 32 apps (negatives gathered over all GPUs) \\
Epochs / steps & 3 / 2,379; fp16 autocast, gradient checkpointing \\
Selection & mean validation AUC of 3 LR probes (static, dynamic, fusion) on 1,000 training apps, twice per epoch \\
Hardware & 8 $\times$ NVIDIA Quadro RTX 4000 (8\,GB); peak 2.8\,GB per GPU; 3.2--3.4\,h per run \\
Downstream LR & standardized inputs; $C \in \{0.01,\dots,1000\}$ by 5-fold CV on training data \\
\bottomrule
\end{tabularx}
\end{table}

\subsection{Using the Shared Space}
\label{sec:use}

We use only simple downstream models, so that the results measure the space and not the classifier.
A probe standardizes its inputs on the training fingerprints of its own view.
\emph{Fusion} uses the concatenation $[z^{s};z^{d}]$.
\emph{Transfer \sd{}} trains the probe on static fingerprints and applies it unchanged to dynamic fingerprints; \emph{transfer \ds{}} is the reverse.
\emph{Matching} uses a report as the query and ranks the static fingerprints of a \emph{gallery} of known APKs by cosine similarity.
\emph{Verification} accepts the claim ``report $r$ comes from APK $a$'' if $z^{d}_r\cdot z^{s}_a > t$.
\emph{Family attribution} gives a report the similarity-weighted family vote of its $k{=}10$ nearest static fingerprints in a labelled gallery.

\section{Experimental Setup}
\label{sec:setup}

\paragraph*{Data}
All apps have both an APK and a DroidBox report from the same DroidBox setup, which MalDy~\cite{karbab2019maldy} also used.
The benign apps come from AndroZoo~\cite{allix2016androzoo}, and the malware comes from MalDozer~\cite{karbab2018maldozer}.
We join reports and APKs by hash, drop four malware apps that fail (corrupt archives or no framework APIs), and sample 18,140 benign apps to balance the 18,136 malware apps, which gives 36,276 apps (\cref{tab:data}).
The bulk of both classes is from 2011--2014 by the build date of \texttt{classes.dex}, but the distributions differ: 2011--2012 holds 44\% of the malware and 28\% of the benign apps, and 14\% of the malware is from 2015--2016, where we have no benign apps.
The labels follow the two sources.
All benign apps and 18,078 of the 18,136 malware apps (99.7\%) are in AndroZoo, so readers can download 36,218 of the 36,276 APKs with the hashes in our repository.

\paragraph*{Splits}
\splitR{} splits the apps at random (70/15/15), stratified by class.
\splitF{} assigns whole malware families to one part: the seven test families (Adwo, InfoStealer, Kmin, MobilePay, SMSReg, Utchi, Youmi; 3,158 apps) and the ten validation families do not occur in training, so \splitF{} measures generalization to unseen families.
Subsets of 4,000 and 16,000 apps (4k, 16k) serve for the search and the ablation.

\paragraph*{Protocol and metrics}
All choices (hyperparameters, checkpoints, probe regularization by 5-fold cross-validation) use the training and validation data only, and we used the test data once for each final model.
For detection and transfer, the metric is the area under the ROC curve (AUC).
For matching, we report the recall at $k$ (\Rk{k}, the share of queries whose APK is in the top $k$), the median rank, and the mean reciprocal rank (MRR); random matching in the 5,442 test apps gives $\Rk{10}=0.002$.
For verification, we report the AUC, the equal error rate (EER) and the true positive rate at 1\% false positive rate.

\paragraph*{Baselines}
The baseline is the same encoder without fine-tuning (\frz{}), with the same instructions, pooling, inputs and downstream models, which isolates the effect of contrastive fine-tuning.
Because \frz{} was never asked to align the views, we also fit a canonical correlation analysis (CCA) on the same training pairs, with its dimension (32--256) chosen by the validation MRR.
A feature-based detector of one view is not a baseline for our questions: it cannot transfer across views, match a report to an APK, or verify a pair.

\paragraph*{Statistics and implementation}
We give 95\% confidence intervals (CI) from a paired, class-stratified bootstrap over the test apps~\cite{efron1993introduction}; a difference is significant when its CI excludes zero.
We trained \detm{} with three seeds on \splitR{} and report the mean $\pm$ standard deviation; the other analyses use seed 0, the run that the search selected.
\Cref{tab:seeds} gives all three seeds; the CIs cover the sampling of the test apps.
\matm{} and the \splitF{} models have one seed each.

\begin{table}[t]
\caption{Three seeds of \detm{} on \splitR{} (5,442 test apps, LR probe; std with ddof 1).}
\label{tab:seeds}
\centering
\footnotesize
\setlength{\tabcolsep}{3pt}
\begin{tabular}{@{}l c c c c@{}}
\toprule
Metric & Seed 0 & Seed 1 & Seed 2 & Mean $\pm$ std \\
\midrule
Static AUC & 0.9948 & 0.9955 & 0.9959 & 0.9954 $\pm$ 0.0005 \\
Dynamic AUC & 0.9781 & 0.9784 & 0.9794 & 0.9786 $\pm$ 0.0007 \\
Fusion AUC & 0.9959 & 0.9972 & 0.9971 & 0.9967 $\pm$ 0.0007 \\
Transfer \sd{} AUC & 0.9761 & 0.9733 & 0.9770 & 0.9755 $\pm$ 0.0019 \\
Transfer \ds{} AUC & 0.9930 & 0.9932 & 0.9933 & 0.9931 $\pm$ 0.0001 \\
Matching \ds{} \Rk{1} & 0.029 & 0.012 & 0.027 & 0.023 $\pm$ 0.009 \\
Matching \ds{} \Rk{10} & 0.137 & 0.089 & 0.110 & 0.112 $\pm$ 0.024 \\
Matching \ds{} MRR & 0.065 & 0.042 & 0.057 & 0.055 $\pm$ 0.012 \\
Verification AUC & 0.871 & 0.852 & 0.860 & 0.861 $\pm$ 0.009 \\
\bottomrule
\end{tabular}
\end{table}
The preprocessing reads the APKs in memory and never runs them.

\section{Results}
\label{sec:results}

\Cref{tab:headline} gives the main test results.
We explain them through five research questions:
\textbf{RQ1} Does fine-tuning align the two views?
\textbf{RQ2} Does a detector transfer across views?
\textbf{RQ3} Is fusion robust when one view is weak?
\textbf{RQ4} Does the model learn behaviour, not shortcuts?
\textbf{RQ5} Can the fingerprints link a report to its APK, verify a pair and attribute a family?
\Cref{tab:headline} shows one pattern: \detm{} is the best model for detection and transfer, and \matm{} is the best model for matching and verification.
Both are above \frz{} in every row, by a small margin for single-view detection and by a large margin for transfer, matching and verification.
The single-view detection AUC alone therefore does not show what the space can do; the cross-view rows do.

\begin{table}[t]
\caption{Main test results (LR probe). \detm{} on \splitR{}: mean $\pm$ std of 3 seeds; other models: one seed. Matching: a report is the query; the gallery holds all test apps (5,442 in \splitR{}, 5,879 in \splitF{}). Verification: same-app pairs against random pairs. Bold: best in the row. Every \detm{}$-$\frz{} (seed 0) and \matm{}$-$\frz{} difference has a 95\% CI above zero.}
\label{tab:headline}
\centering
\footnotesize
\setlength{\tabcolsep}{3pt}
\begin{tabular}{@{}l c c c@{}}
\toprule
Task & \detm{} & \matm{} & \frz{} \\
\midrule
\multicolumn{4}{@{}l}{\emph{\splitR{}}} \\
Detection AUC, static   & \textbf{0.9954} $\pm$ 0.0005 & 0.993 & 0.978 \\
Detection AUC, dynamic  & \textbf{0.9786} $\pm$ 0.0007 & 0.971 & 0.962 \\
Detection AUC, fusion   & \textbf{0.9967} $\pm$ 0.0007 & 0.995 & 0.989 \\
Transfer AUC, \sd{}     & \textbf{0.975} $\pm$ 0.002 & 0.920 & 0.736 \\
Transfer AUC, \ds{}     & \textbf{0.993} $\pm$ 0.000 & 0.977 & 0.625 \\
Matching \ds{} \Rk{1}   & 0.023 $\pm$ 0.009 & \textbf{0.141} & 0.001 \\
Matching \ds{} \Rk{10}  & 0.112 $\pm$ 0.024 & \textbf{0.466} & 0.007 \\
Matching \ds{} MRR      & 0.055 $\pm$ 0.012 & \textbf{0.244} & 0.004 \\
Median rank of the APK  & 207 (seed 0) & \textbf{13} & 2,714 \\
Verification AUC        & 0.861 $\pm$ 0.009 & \textbf{0.983} & 0.512 \\
\midrule
\multicolumn{4}{@{}l}{\emph{\splitF{} (unseen test families)}} \\
Detection AUC, static   & \textbf{0.975} & 0.944 & 0.929 \\
Detection AUC, dynamic  & 0.924 & \textbf{0.928} & 0.882 \\
Detection AUC, fusion   & \textbf{0.984} & 0.966 & 0.948 \\
Transfer AUC, \sd{}     & \textbf{0.913} & 0.875 & 0.639 \\
Transfer AUC, \ds{}     & \textbf{0.974} & 0.907 & 0.604 \\
Matching \ds{} \Rk{10}  & 0.032 & \textbf{0.178} & 0.005 \\
Verification AUC        & 0.775 & \textbf{0.925} & 0.507 \\
\bottomrule
\end{tabular}
\end{table}

\subsection{RQ1: Does Fine-Tuning Align the Two Views?}
\label{sec:rq1}

\begin{figure}[t]
\centering
\includegraphics[width=\linewidth]{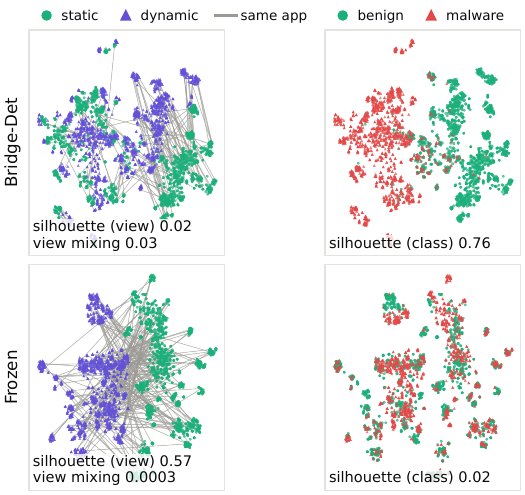}
\caption{t-SNE maps of 2,000 test apps (\splitR{}), both views. Left: coloured by view; grey lines join the two views of 200 apps. Right: coloured by class. Top: \detm{}. Bottom: \frz{}. In the frozen space, the views form two regions; after fine-tuning, the views share the class regions.}
\label{fig:maps}
\end{figure}

\paragraph*{The frozen space separates the views, not the apps}
\Cref{fig:maps} shows t-SNE maps~\cite{vandermaaten2008visualizing} of 2,000 test apps with both views.
In the \frz{} space, the static and dynamic fingerprints form two separate regions (silhouette by view 0.57), and among the ten nearest neighbours of a static fingerprint, only 0.035\% are dynamic fingerprints (view mixing 0.00035; a fully mixed space gives 0.5).
Benign apps and malware are mixed (silhouette by class 0.02).
The pretrained encoder thus sees mostly the \emph{form} of the text: a list of API names looks different from a list of events, whatever the app does.

\paragraph*{The fine-tuned space shares the class regions}
In the \detm{} space, the two views fall into the same class regions (silhouette by view 0.02), and the classes separate (silhouette by class 0.76); all silhouettes come from the full 1,024-dimensional space.
The class silhouette is a direct target of SupCon, so it shows class structure, not alignment; matching is the direct test of alignment.

\begin{figure}[t]
\centering
\includegraphics[width=\linewidth]{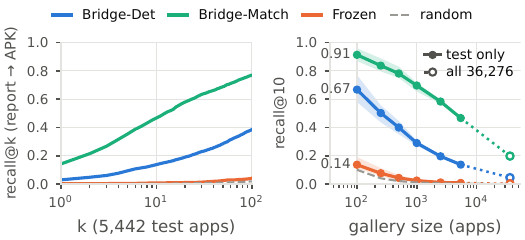}
\caption{Matching (report to APK) on \splitR{}. Left: recall at $k$ in the gallery of 5,442 test apps. Right: \Rk{10} against the gallery size (mean and min--max of 20 random galleries); the dotted end point uses all 36,276 apps. \detm{} is seed 0.}
\label{fig:matching}
\end{figure}

\paragraph*{Matching quantifies the alignment}
A report is the query, and we rank the static fingerprints of a gallery.
\frz{} is near chance (\cref{fig:matching}): its \Rk{10} is 0.007 in the test gallery, and the median rank of the true APK is 2,714, the middle of the gallery.
With a gallery of 100 apps, \detm{} (seed 0) puts the true APK in the top 10 for 67\% of the reports and \matm{} for 91\% (\frz{}: 14\%); with 1,000 apps, the values are 29\% and 70\% (\frz{}: 2\%).
With all 36,276 apps as the gallery (training apps included), \matm{} still finds the true APK in the top 10 for 20\% of the test reports.

\paragraph*{A linear alignment of the frozen space}
\Cref{tab:aligned} compares CCA, fitted on the same training pairs, with the fine-tuned models.
CCA recovers much of the cross-view ability: on \splitR{}, it transfers with an AUC of 0.934 (\sd{}) and 0.955 (\ds{}) and verifies pairs with 0.905.
The frozen space thus already contains much of the information that links the views, but in different directions for each view.
Contrastive fine-tuning still gives the best transfer (\detm{}: +0.042 and +0.038 over CCA, CIs excluding zero) and the best verification (\matm{}: 0.983 against 0.905), and the gaps grow on unseen families (transfer 0.913 and 0.974 against 0.864 and 0.884; verification 0.925 against 0.826).
For matching alone, CCA is as good as \matm{}: its \Rk{1} and MRR are higher (0.224 and 0.297 against 0.141 and 0.244), and its \Rk{10} is close (0.448 against 0.466, overlapping CIs).

\begin{table}[t]
\caption{CCA (256 dim.) of the frozen fingerprints, fitted on the training pairs, against the fine-tuned models (test; \detm{} seed 0). \detm{}: seed 0. This analysis fits its own probes and draws new verification negatives, so values can differ by up to 0.002 from \cref{tab:headline}.}
\label{tab:aligned}
\centering
\footnotesize
\setlength{\tabcolsep}{3pt}
\begin{tabular}{@{}l c c c c c@{}}
\toprule
Model & \sd{} & \ds{} & \Rk{1} & \Rk{10} & Verif. \\
\midrule
\multicolumn{6}{@{}l}{\emph{\splitR{}}} \\
\frz{}, raw          & 0.736 & 0.625 & 0.001 & 0.007 & 0.512 \\
\frz{} + CCA         & 0.934 & 0.955 & \textbf{0.224} & 0.448 & 0.905 \\
\detm{}              & \textbf{0.976} & \textbf{0.993} & 0.029 & 0.137 & 0.871 \\
\matm{}              & 0.920 & 0.977 & 0.141 & \textbf{0.466} & \textbf{0.983} \\
\multicolumn{6}{@{}l}{\emph{\splitF{}}} \\
\frz{}, raw          & 0.640 & 0.605 & 0.001 & 0.005 & 0.509 \\
\frz{} + CCA         & 0.864 & 0.884 & \textbf{0.068} & 0.162 & 0.826 \\
\detm{}              & \textbf{0.913} & \textbf{0.974} & 0.007 & 0.032 & 0.773 \\
\matm{}              & 0.875 & 0.907 & 0.049 & \textbf{0.178} & \textbf{0.925} \\
\bottomrule
\end{tabular}
\end{table}

\finding{1} The raw frozen encoder does not align the two views: it separates the apps by view, and its matching is near chance. A linear CCA recovers much of the cross-view ability; contrastive fine-tuning gives the best transfer and the best verification, with the largest margin on unseen families.

\subsection{RQ2: Does a Detector Transfer Across Views?}
\label{sec:rq2}

We train an LR on one view of the training apps and test it, unchanged, on the other view of the test apps (S1, S2).
On \splitR{}, \detm{} transfers with an AUC of $0.975\pm0.002$ (\sd{}) and $0.993\pm0.000$ (\ds{}), against 0.736 and 0.625 for \frz{} (\cref{tab:headline}).
The transfer AUC is bounded by the separability of the target view, and it is close to the in-view AUC of that view: 0.993 against 0.995 on APKs, and 0.975 against 0.979 on reports.
On \splitF{}, the transfer AUCs are 0.913 and 0.974, against 0.639 and 0.604 for \frz{}: the transfer stays above 0.91 on malware families that the model has never seen.

\paragraph*{Transfer without labels}
The cross-view SupCon term of \detm{} pulls static malware fingerprints towards dynamic malware fingerprints, so part of its transfer is a property that training asked for.
\matm{} is the cleaner test of scenario S1/S2, because its encoder never saw a label: it only learned which report belongs to which APK.
Its probe transfers with 0.920 (\sd{}) and 0.977 (\ds{}) on \splitR{} (0.955 and 0.987 with the classifier chosen on the validation data), and with 0.875 and 0.907 on \splitF{}.
The alignment of the two views alone thus gives most of the transfer; the SupCon terms add the rest.

\begin{figure}[t]
\centering
\includegraphics[width=\linewidth]{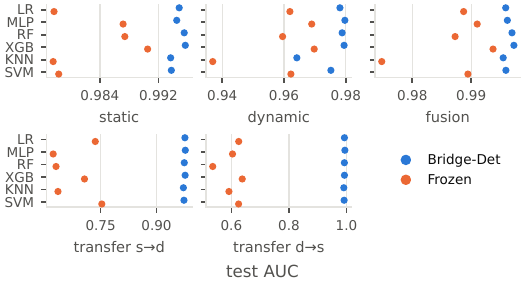}
\caption{Test AUC of six classifiers on five inputs (\splitR{}), each chosen on the validation data. On \detm{} fingerprints, the classifiers agree within 0.005 AUC for static detection and fusion. The frozen space does not transfer with any classifier.}
\label{fig:classifiers}
\end{figure}

\paragraph*{The result does not depend on the classifier}
With six classifiers (LR, MLP, random forest, XGBoost, $k$-NN, linear SVM) chosen on the validation data, all classifiers are within 0.005 AUC of each other for static detection and fusion on \detm{} fingerprints (\cref{fig:classifiers}; Supplementary Fig.~\ref{fig:classifiers_family} for \splitF{}).
With the best classifier, \frz{} transfer stays at 0.753 (\sd{}) and 0.625 (\ds{}), against 0.977 and 0.995 for \detm{} (seed 0): transfer needs an alignment, not a better decision function.

\paragraph*{Family knowledge also transfers}
A $k$-NN family classifier trained on one view of the 31 families with at least 20 training apps and tested on the other view gives a macro-F1 of 0.35 (\sd{}) and 0.40 (\ds{}) with \detm{}, against 0.03 and 0.02 with \frz{}.
Within one view, the two models are almost the same (LR macro-F1 for static, dynamic and fusion: \detm{} 0.875, 0.768, 0.891; \frz{} 0.857, 0.785, 0.882).
Fine-tuning thus does not add family information inside a view; it makes the family information of one view usable in the other view.

\finding{2} A classifier trained on one view detects malware from the other view with an AUC of 0.975 (\sd{}) and 0.993 (\ds{}), and of 0.913 and 0.974 on unseen families. \matm{}, which never saw a label, also transfers (0.920 and 0.977). The frozen encoder does not transfer with any classifier (0.60--0.75).

\subsection{RQ3: Is Fusion Robust When One View Is Weak?}
\label{sec:rq3}

All \detm{} values in this subsection are for seed 0 on \splitR{}.

\paragraph*{Error accounting}
At a threshold of 0.5, the dynamic probe misclassifies 359 test apps (6.6\%); fusion corrects 297 of them and adds 36 new errors.
The static probe misclassifies 109 apps (2.0\%); fusion corrects 33 and adds 22.
Fusion therefore helps mostly the weaker dynamic view.
On \splitF{}, the threshold of the fusion probe transfers less well: at 0.5, fusion makes 74 more errors than the static probe.
The fusion error rate is 1.8\% for \detm{} and 4.3\% for \frz{} on \splitR{}, and 11.7\% and 18.4\% on \splitF{}.

\paragraph*{Natural weak views}
\Cref{tab:weak} gives the AUC on test subsets in which one view is naturally weak.
Reports with 0 to 10 events give a dynamic AUC of only 0.894, but fusion keeps 0.991.
Apps with a DEX file in \texttt{assets/} (92\% malware) stand for packed loaders, whose real code may be outside the analysed DEX files; their static AUC falls to 0.984, and fusion gives 0.991.
In every subset, the \detm{} fusion AUC is 0.991 or higher and above the \frz{} fusion.

\begin{table}[t]
\caption{Detection AUC on test subsets with a naturally weak view (\splitR{}, LR probe, \detm{} seed 0). Low API: fewer than 50 external APIs. Budget cut: the static list was cut to 2,048 tokens.}
\label{tab:weak}
\centering
\footnotesize
\setlength{\tabcolsep}{2.4pt}
\begin{tabular}{@{}l r r c c c c@{}}
\toprule
 & & & \multicolumn{3}{c}{\detm{}} & \frz{} \\
\cmidrule(lr){4-6}\cmidrule(l){7-7}
Subset & Apps & Mal. & Static & Dyn. & Fusion & Fusion \\
\midrule
All test apps        & 5,442 & 2,721 & 0.995 & 0.978 & 0.996 & 0.989 \\
0--10 events         &   879 &   354 & 0.994 & 0.894 & 0.991 & 0.976 \\
11--50 events        & 3,464 & 1,604 & 0.995 & 0.979 & 0.996 & 0.989 \\
$>$50 events         & 1,099 &   763 & 0.993 & 0.998 & 0.999 & 0.994 \\
DEX in \texttt{assets/} & 306 &   281 & 0.984 & 0.974 & 0.991 & 0.966 \\
Low API ($<$50)      &   347 &   239 & 0.999 & 0.922 & 0.999 & 0.996 \\
Budget cut           & 3,495 & 1,425 & 0.993 & 0.980 & 0.995 & 0.984 \\
\bottomrule
\end{tabular}
\end{table}

\begin{figure}[t]
\centering
\includegraphics[width=\linewidth]{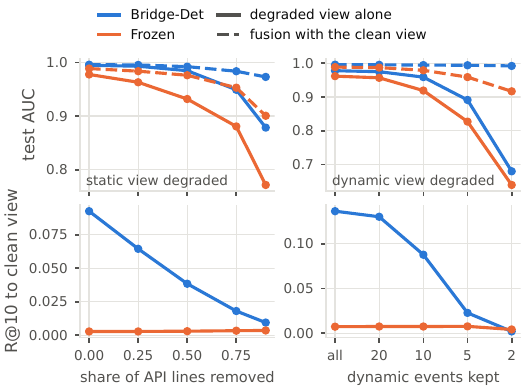}
\caption{Controlled degradation of one test view (\splitR{}, \detm{} seed 0). Top: detection AUC of the degraded view alone (solid) and of the fusion with the clean other view (dashed). Bottom: matching \Rk{10} of the degraded view against the clean other view. Probes are trained on clean data.}
\label{fig:degradation}
\end{figure}

\paragraph*{Controlled degradation}
We degrade one view of the test apps and keep the other view clean: we remove a random share (25--90\%) of the API lines, or we keep only $n$ random events ($n=20, 10, 5, 2$).
The probes are trained on clean data, so the model has never seen degraded inputs.
\Cref{fig:degradation} shows that the degraded view alone loses much of its power, but the \detm{} fusion loses little (Supplementary Table~\ref{tab:degrade} gives the values).
With 90\% of the API lines removed, the static AUC falls to 0.879, and with two events, the dynamic AUC falls to 0.680.
Fusion with the clean view keeps 0.973 and 0.992, within 0.005 of the clean view alone (0.978 and 0.995).
\frz{} fusion falls to 0.901 and 0.917, 0.061 below its clean view: in the frozen space, a weak view harms the fusion.
The degraded fingerprint stays close to the clean one: with half of the API lines removed, its cosine similarity to the clean static fingerprint is 0.98.
Matching is more sensitive than detection, because it needs the details of an app and not only its class: with half of the API lines removed, the \Rk{10} of matching with the APK as the query falls from 0.093 to 0.038.

\finding{3} Under random degradation, when one view is almost empty (two events, or 10\% of the APIs), the \detm{} fusion keeps an AUC of 0.992 and 0.973, within 0.005 of the clean view alone; \frz{} fusion falls to 0.917 and 0.901. Fusion corrects most errors of the weaker dynamic view.

\subsection{RQ4: Does the Model Learn Behaviour?}
\label{sec:rq4}

The benign apps and the malware come from different collections, and their year distributions differ.
A model can therefore separate the classes by features that are not behaviour~\cite{arp2022dos,pendlebury2019tesseract}.
\Cref{tab:shortcuts} collects our controls (\detm{} seed 0); no control subset lowers the static AUC below 0.987.

\paragraph*{Input length}
Benign apps have about twice as many external APIs as malware (median 1,106 against 597).
If the model learned ``long list = benign'', its AUC would fall inside groups of apps with the same input length.
Over five length quintiles of the probed input, the mean AUC is 0.994 (static) and 0.975 (dynamic), the same as on all apps.

\paragraph*{Time period}
With benign apps up to 2012 (787 apps) or from 2013 (1,934 apps) against all test malware, \detm{} gives a static AUC of 0.994 and 0.995 and a dynamic AUC of 0.969 and 0.982.
The age of the benign apps thus does not decide the static result.
A probe on benign apps predicts the year group from the static fingerprint with an AUC of 0.93, and the \frz{} fingerprint gives almost the same value (0.92): fine-tuning neither adds nor removes this information.

\paragraph*{Metadata and duplicates}
Eight metadata features (API count, token counts, cut flag, event count, low-API and \texttt{assets/} flags) give an AUC of 0.930 with XGBoost on \splitR{} and 0.795 on \splitF{}.
\detm{} is clearly above the best metadata model: +0.065 [+0.058, +0.071] for static detection on \splitR{} and +0.180 [+0.167, +0.191] on \splitF{}.
Repackaged apps and versions put almost the same app into training and test data: 47.8\% of the \splitR{} test apps have a static text identical to a training app.
On the 2,842 test apps without such a duplicate, \detm{} gives a static AUC of 0.987 and transfers with 0.965 (\sd{}) and 0.985 (\ds{}), 0.008--0.011 below all test apps.
By construction, \splitF{} has no duplicates of the test families in training.

\begin{table}[t]
\caption{Shortcut controls (test, \detm{} seed 0, LR probe, \splitR{} unless stated). Leakage probe: benign apps only; the probe predicts the year group from the fingerprint (\detm{} / \frz{}).}
\label{tab:shortcuts}
\centering
\footnotesize
\setlength{\tabcolsep}{2.6pt}
\begin{tabular}{@{}l r c c c@{}}
\toprule
Control & Apps & Static & Dyn. & Fusion \\
\midrule
All test apps                         & 5,442 & 0.995 & 0.978 & 0.996 \\
No static duplicate in training       & 2,842 & 0.987 & 0.966 & 0.989 \\
Benign up to 2012 vs.\ malware        & 3,508 & 0.994 & 0.969 & 0.995 \\
Benign from 2013 vs.\ malware         & 4,655 & 0.995 & 0.982 & 0.996 \\
Benign 2011--2014, all malware        & 5,393 & 0.995 & 0.978 & 0.996 \\
Length-matched (mean of 5 quintiles)  & 5,442 & 0.994 & 0.975 & 0.995 \\
\midrule
Leakage probe: year group ($\geq$2013) & 2,721 & 0.93\,/\,0.92 & 0.80\,/\,0.80 & -- \\
\midrule
\multicolumn{5}{@{}l}{\emph{Metadata only (8 features, no fingerprint), LR / XGBoost}} \\
\splitR{}                             & 5,442 & \multicolumn{3}{c}{0.818 / 0.930} \\
\splitF{}                             & 5,879 & \multicolumn{3}{c}{0.712 / 0.795} \\
\bottomrule
\end{tabular}
\end{table}

\finding{4} The AUC stays the same inside each length quintile and each benign age group, and it drops by at most 0.012 without exact duplicates. The fingerprints are clearly above a metadata-only model (AUC 0.930): +0.065, and +0.180 on unseen families. The time controls vary the years of the benign apps.

\subsection{RQ5: Cross-View Forensic Linking}
\label{sec:rq5}

This subsection tests the space as a source of forensic fingerprints for individual apps.
In scenario S4, the analyst has a sandbox report and wants to know which known APK produced it (matching) or whether a given APK produced it (verification).
In scenario S2, the analyst wants to attribute a report to a family.

\begin{figure}[t]
\centering
\includegraphics[width=\linewidth]{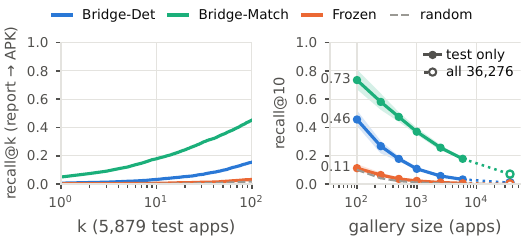}
\caption{Matching (report to APK) on \splitF{} (gallery of 5,879 test apps, 3,158 of them malware from unseen families). Same layout as \cref{fig:matching}.}
\label{fig:matching_family}
\end{figure}

\paragraph*{The objective decides the matching quality}
\matm{} finds the true APK of a report in the top 10 of 5,442 test apps for 46.6\% of the reports, against $11.2\pm2.4$\% for \detm{} (\cref{tab:headline,fig:matching}).
Its median rank of the true APK is 13, against 207 for \detm{} and 2,714 for \frz{}.
The paired CIs of \matm{}$-$\detm{} (seed 0) exclude zero: +0.113 [+0.103, +0.123] for \Rk{1}, +0.330 [+0.316, +0.344] for \Rk{10} and +0.179 [+0.170, +0.188] for MRR.
As \cref{sec:loss} explains, SupCon removes the differences between apps of one class, so in a gallery many malware apps look alike; InfoNCE alone keeps each app distinct.
On \splitF{}, \matm{} reaches \Rk{10} 0.178 (\detm{} 0.032), and 0.092 on the test malware of unseen families alone (\detm{} 0.021, \frz{} 0.004); \cref{fig:matching_family} shows the curves.

\begin{figure}[t]
\centering
\includegraphics[width=\linewidth]{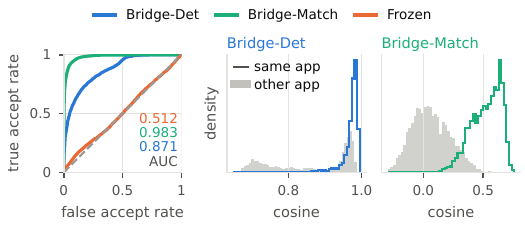}
\caption{Report--APK verification on \splitR{}. Left: ROC of the cosine similarity for pairs of the same app against random pairs. Right: distribution of the similarity for the two kinds of pairs. \detm{} is seed 0.}
\label{fig:verification}
\end{figure}

\paragraph*{Verification}
The score of a report--APK pair is the cosine similarity of the two fingerprints.
Against random pairs of different apps, \matm{} separates true pairs with an AUC of 0.983 [0.981, 0.985], an EER of 0.066 and a true positive rate of 0.73 at a false positive rate of 1\% (\cref{fig:verification}).
\detm{} gives $0.861\pm0.009$ (seed 0: EER 0.22, TPR 0.26 at 1\% FPR), and \frz{} is at chance (AUC 0.512, EER 0.50).
Random pairs are easy negatives, because most of them mix a benign app and a malware app.
\Cref{tab:hardneg} therefore also uses harder negatives.
\matm{} stays strong against different apps of the same class (AUC 0.972) and loses more against different apps of the same family (0.885); \detm{} falls to 0.759 and 0.717.
Same-family pairs are the realistic hard case for forensics, because the apps of one family often share most of their code.

\begin{table}[t]
\caption{Verification (Verif.) AUC with negatives of increasing difficulty, and matching \Rk{10} with gallery and queries of one class only (test; \detm{} seed 0; new draw of negatives).}
\label{tab:hardneg}
\centering
\footnotesize
\setlength{\tabcolsep}{3pt}
\begin{tabular}{@{}l c c c@{}}
\toprule
Metric & \detm{} & \matm{} & \frz{} \\
\midrule
\multicolumn{4}{@{}l}{\emph{\splitR{}}} \\
Verif., random negatives      & 0.871 & \textbf{0.983} & 0.512 \\
Verif., same-class negatives  & 0.759 & \textbf{0.972} & 0.509 \\
Verif., same-family negatives & 0.717 & \textbf{0.885} & 0.499 \\
Match \Rk{10}, benign only       & 0.054 & \textbf{0.343} & 0.010 \\
Match \Rk{10}, malware only      & 0.225 & \textbf{0.622} & 0.010 \\
\multicolumn{4}{@{}l}{\emph{\splitF{}}} \\
Verif., random negatives      & 0.773 & \textbf{0.923} & 0.509 \\
Verif., same-family negatives & 0.584 & \textbf{0.748} & 0.504 \\
Match \Rk{10}, malware only      & 0.023 & \textbf{0.107} & 0.006 \\
\bottomrule
\end{tabular}
\end{table}

\paragraph*{Matching per class and without duplicates}
Matching is easier for malware than for benign apps (\cref{tab:hardneg}): many malware apps have distinctive reports (premium numbers, command-and-control hosts), while many benign reports show only generic file and network activity.
On the 2,842 test apps without a static duplicate in training, \matm{} reaches \Rk{10} 0.393, against 0.564 on random subsets of the same size; this subset has more benign apps (69\% against 50\%), which explains part of the drop.

\paragraph*{The cost of InfoNCE only}
Compared with \detm{}, the static and dynamic detection AUCs of \matm{} are 0.002 and 0.007 lower on \splitR{}, and the fusion difference is not significant.
The transfer cost is larger: $-0.056$ (\sd{}) and $-0.016$ (\ds{}) on \splitR{}, and $-0.038$ and $-0.067$ on \splitF{}.
The SupCon terms thus matter most where the model must generalize: across views and to unseen families.
We therefore keep both models; the second model costs one more training run.

\begin{figure}[t]
\centering
\includegraphics[width=\linewidth]{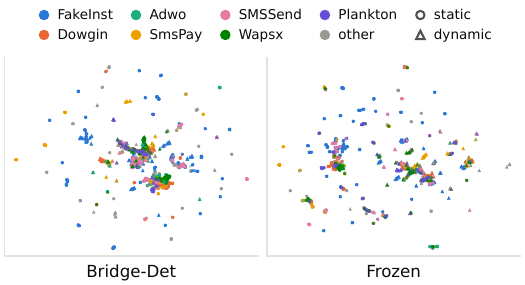}
\caption{UMAP map~\cite{mcinnes2018umap} of the malware test apps (\splitR{}), coloured by the seven largest families, both views. With \detm{}, the two views of a family fall into the same regions; with \frz{}, they do not.}
\label{fig:familymap}
\end{figure}

\paragraph*{Cross-view family attribution and clustering}
A malware report gets the family vote of its 10 nearest static fingerprints in a labelled gallery of training malware.
\detm{} gives a top-1 accuracy of 0.534, against 0.187 for \frz{} and 0.27 for the majority class (within the static view: 0.864 and 0.822).
\Cref{fig:familymap} shows that with \detm{}, the static and dynamic fingerprints of a family fall into the same regions, and with \frz{}, they do not.
Without labels, agglomerative clustering of the static fingerprints of the test malware gives an adjusted Rand index of 0.256 and a normalized mutual information of 0.449 with the families (\frz{}: 0.024 and 0.089).

\begin{figure}[t]
\centering
\includegraphics[width=\linewidth]{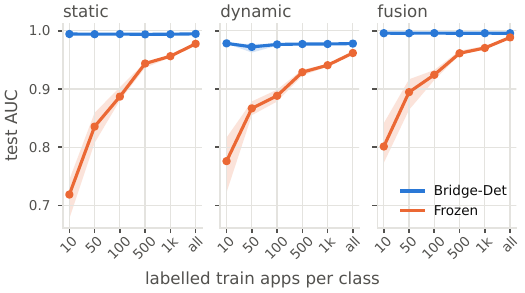}
\caption{Label efficiency (\splitR{}): test AUC of LR probes trained with $n$ labelled apps per class (mean of 5 draws). The encoder of \detm{} was trained with all training labels; the figure measures how easily a new probe learns from the space.}
\label{fig:labels}
\end{figure}

\begin{figure}[t]
\centering
\includegraphics[width=\linewidth]{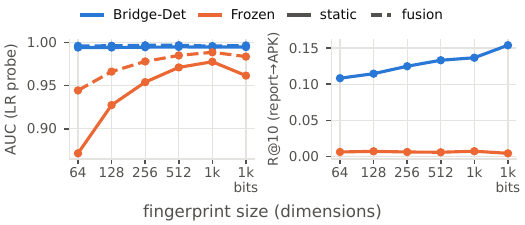}
\caption{Compact fingerprints (\splitR{}). Detection AUC (left) and matching \Rk{10} (right) against the fingerprint size, from 64 float dimensions to 1,024 sign bits (1k = 1,024). \detm{} keeps its AUC down to 64 dimensions.}
\label{fig:compact}
\end{figure}

\paragraph*{Compact fingerprints}
A fingerprint database for a large collection must be small and fast to search.
The base model is trained with Matryoshka representation learning~\cite{kusupati2022matryoshka}, so we can truncate the fingerprint to its first dimensions.
With 64 dimensions (16 times smaller), the \detm{} static detection AUC is 0.994 and the matching \Rk{10} is 0.108; \frz{} gives an AUC of 0.871.
With 1,024 sign bits (128 bytes, 32 times smaller than float32), the AUC is 0.995 and \Rk{10} is 0.154, so binarization reduces neither value and allows search with the Hamming distance (\cref{fig:compact}; Supplementary Fig.~\ref{fig:compact_family} for \splitF{}).
A new probe on this space also needs few labels: with 10 labelled apps per class, the static AUC is 0.995, against 0.72 for \frz{} (\cref{fig:labels}; the encoder itself used all training labels).

\paragraph*{Toward open-set attribution}
Unseen families are detected as \emph{malware} (\splitF{} AUC 0.975), and their fingerprints lie close to the known families: scored by the cosine distance to the nearest family centroid, they reach a static AUC of 0.620 (\detm{}), 0.635 (\matm{}) and 0.622 (\frz{}) against held-out training malware (10\%) of known families.
Open-set attribution is therefore a natural extension of the space.

\finding{5} With the right objective, the space links individual apps across views. \matm{} ranks the true APK of a report at median 13 of 5,442 apps and verifies report--APK pairs with an AUC of 0.983 (0.885 against same-family pairs). \detm{} attributes reports to families across views (0.534 top-1, majority 0.27), and its fingerprints stay useful at 64 dimensions or as 128-byte bit strings.

\section{Ablation and Data Size}
\label{sec:ablation}

We ran a leave-one-out ablation of the final \detm{} configuration on the 16k subset (2,400 test apps), with one setting changed per run (\cref{tab:ablation}; Supplementary Fig.~\ref{fig:ablation}).

\begin{table}[t]
\caption{Leave-one-out ablation on the 16k subset (\splitR{} test, 2,400 apps, LR probe). Effect = run $-$ final configuration. Bold: the 95\% CI excludes zero. n/a: no training embeddings stored.}
\label{tab:ablation}
\centering
\scriptsize
\setlength{\tabcolsep}{1.6pt}
\begin{tabular}{@{}l c c c c c c@{}}
\toprule
 & \multicolumn{3}{c}{Detection} & \multicolumn{2}{c}{Transfer} & Match \\
\cmidrule(lr){2-4}\cmidrule(lr){5-6}\cmidrule(l){7-7}
Run & Static & Dyn. & Fusion & \sd{} & \ds{} & \Rk{10} \\
\midrule
Final (absolute)              & 0.995 & 0.973 & 0.996 & 0.972 & 0.995 & 0.038 \\
\midrule
\frz{} (no fine-tuning)       & $\mathbf{-0.025}$ & $\mathbf{-0.012}$ & $\mathbf{-0.010}$ & n/a & n/a & $\mathbf{-0.029}$ \\
Last-token pooling            & $\mathbf{-0.003}$ & $+0.001$ & $-0.001$ & $+0.002$ & $-0.001$ & $\mathbf{+0.028}$ \\
lr $10^{-4}$                  & $\mathbf{-0.002}$ & $+0.002$ & $\mathbf{-0.002}$ & $-0.001$ & $-0.001$ & $-0.008$ \\
2 epochs                      & $-0.001$ & $+0.000$ & $-0.000$ & n/a & n/a & $\mathbf{+0.012}$ \\
No static SupCon              & $-0.001$ & $+0.002$ & $-0.001$ & $-0.001$ & $\mathbf{-0.005}$ & $\mathbf{+0.134}$ \\
No dynamic SupCon             & $-0.002$ & $-0.000$ & $-0.000$ & $-0.000$ & $-0.001$ & $\mathbf{+0.086}$ \\
SupCon only                   & $-0.001$ & $+0.001$ & $-0.000$ & $+0.001$ & $-0.001$ & $\mathbf{-0.023}$ \\
InfoNCE only                  & $-0.002$ & $-0.002$ & $-0.001$ & $\mathbf{-0.043}$ & $\mathbf{-0.029}$ & $\mathbf{+0.476}$ \\
\bottomrule
\end{tabular}
\end{table}

\paragraph*{Fine-tuning is the largest effect}
Without fine-tuning, every metric falls significantly; the static AUC falls by 0.025.
Each other setting changes the detection AUC by at most 0.003.
The three seeds of the final configuration have a standard deviation of 0.0005--0.0007 on the detection AUCs, so single-seed differences below about 0.002 are within the training variance, even when their CI excludes zero.

\begin{figure}[t]
\centering
\includegraphics[width=\linewidth]{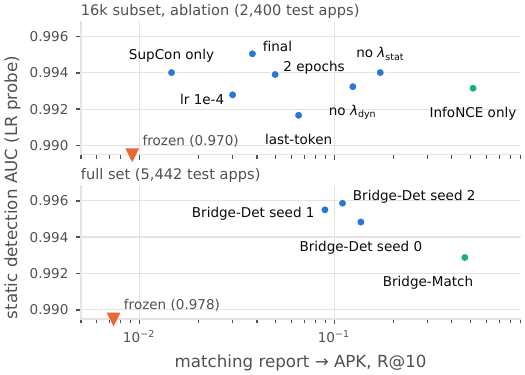}
\caption{Detection against matching. Each point is one model; x: \ds{} \Rk{10} (log scale); y: static detection AUC. Left: ablation runs on the 16k subset. Right: models on the full set (\detm{} seeds 0--2, \matm{}). No $\lambda_{\mathrm{stat}}$/$\lambda_{\mathrm{dyn}}$: no static/dynamic SupCon. \frz{} is below the axis range; its value is at the marker.}
\label{fig:tradeoff}
\end{figure}

\paragraph*{The loss terms trade class structure against app identity}
Removing all SupCon terms raises \Rk{10} from 0.038 to 0.514 (+0.476), and removing the static SupCon term alone raises it by 0.134.
These runs do not lose a significant amount of detection AUC, but the InfoNCE-only run loses transfer ($-0.043$ \sd{} and $-0.029$ \ds{}), as on the full set.
Removing InfoNCE lowers the matching further.
\Cref{fig:tradeoff} plots each model in the plane of matching and detection: the InfoNCE-only runs lie far to the right, and the SupCon runs reach the highest detection AUCs.

\paragraph*{Data size}
Trained on 4,000, 16,000 and 36,276 apps, the final configuration gives a static AUC of 0.969, 0.995 and 0.995 and a dynamic AUC of 0.956, 0.973 and 0.978 (Supplementary Fig.~\ref{fig:datasize}).
The full set also reaches a higher \Rk{10} (0.137 for seed 0 in a gallery of 5,442 apps) than the 16k subset (0.038 in a gallery of 2,400 apps), although a larger gallery makes matching harder.
The test sets of the builds differ, but together with the search, these results suggest that the amount of training data is the most important factor after fine-tuning, in particular for the dynamic view.

\section{Discussion}
\label{sec:discussion}

\paragraph*{A forensic workflow on one fingerprint index}
The results suggest a simple deployment.
An analysis pipeline computes the static fingerprint of every incoming APK and the dynamic fingerprint of every sandbox report, and stores both in one index.
One LR probe, trained on any labelled view, scores all fingerprints, whichever view is available (S1, S2), and the fusion probe gives the most robust score when both views exist (S3).
For forensic questions (S4), the index also holds the \matm{} fingerprints.
When a report arrives without a usable hash, the analyst queries the index and gets a short list of candidate APKs: in a gallery of 1,000 apps, the true APK is in the top 10 for 70\% of the reports.
When a feed claims that a report comes from a given APK, the cosine similarity of the two fingerprints verifies the claim, with a TPR of 0.73 at 1\% FPR against random pairs.
This can flag wrong links in shared threat intelligence and complements hash-based identity.
The same index attributes a report to a family through its nearest static fingerprints, so family knowledge from APK analysis serves the triage of reports.
Each binary \detm{} fingerprint needs 128 bytes, so an index of millions of apps fits into memory, and the encoder (0.6 billion parameters plus 10 million LoRA parameters) runs on a single consumer GPU.

\paragraph*{Future directions}
A time-aware study on recent apps with a current sandbox and a realistic class ratio~\cite{pendlebury2019tesseract,kamol2026mcndroid} will show how the space evolves under drift; the family-disjoint split and the deduplicated subset already give conservative estimates.
For forensic use, open galleries with the rejection of reports whose APK is unknown, forged or repackaged links, and adaptive attacks are natural next tests; because the space links the two views, an attack must be consistent in both views to evade the fusion probe.
Drift detectors such as CADE~\cite{yang2021cade} can add open-set attribution.

\section{Conclusion}
\label{sec:conclusion}

We showed that one contrastive text-embedding space can serve as a bridge between static and dynamic Android malware analysis.
The raw frozen encoder separates the two views and does not link them.
After LoRA fine-tuning with a cross-view InfoNCE loss, the static fingerprint of an app and the dynamic fingerprint of its sandbox report lie close together.
In this space, a detector trained on one view scores the other view (AUC 0.975 and 0.993), the fusion stays strong when one view is almost empty, and a report finds and verifies the APK of its app among thousands of candidates.
The choice of the loss decides the use: supervised contrastive terms give class structure for detection and transfer, and InfoNCE alone keeps the identity of individual apps for forensic matching and verification.
To the best of our knowledge, \name{} is the first method that links sandbox reports and APKs in one space. Its cross-view fingerprints give analysts a new forensic capability: to link evidence from the sandbox and from the APK when hashes cannot.

\section*{Code and Data Availability}
The code, the configurations and the dataset of this paper are shared at \codeurl{}. The dataset gives the hashes, AndroZoo metadata, labels, families and splits of all 36,276 apps, the derived text inputs, the trained adapters and the test fingerprints (release v1.0, with SHA-256 checksums).


\clearpage
\onecolumn
\setcounter{section}{0}\setcounter{figure}{0}\setcounter{table}{0}\setcounter{algorithm}{0}
\renewcommand{\thefigure}{S\arabic{figure}}
\renewcommand{\thetable}{S\arabic{table}}
\renewcommand{\thealgorithm}{S\arabic{algorithm}}
\renewcommand{\thesection}{S\arabic{section}}
\def\theHsection{S\arabic{section}}\def\theHfigure{S\arabic{figure}}\def\theHtable{S\arabic{table}}\def\theHalgorithm{S\arabic{algorithm}}
\makeatletter\def\theHALG@line{S\arabic{algorithm}.\arabic{ALG@line}}\makeatother
\section*{Supplementary Material}

This supplement gives the pseudocode, the training settings, the hyperparameter search and the additional results that the paper refers to.
References to sections, equations, tables and figures without the prefix ``S'' are to the paper.

\section{Method details}
\label{app:method}

\begin{figure}[!htbp]
\centering
\includegraphics[width=0.85\linewidth]{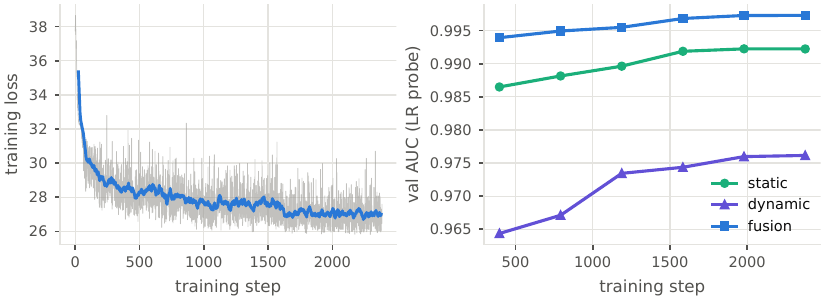}
\caption{Training of the final \detm{} run on \splitR{}: training loss (moving mean) and validation AUC of the selection probes.}
\label{fig:training}
\end{figure}

Algorithm~\ref{alg:inputs} builds the two text views of an app (Section~\ref{sec:inputs} of the paper).
Algorithm~\ref{alg:train} gives one training step, and Algorithm~\ref{alg:use} lists the uses of the fingerprints.
Table~\ref{tab:hparams} of the paper lists all training settings; the two models differ only in the loss weights.
Table~\ref{tab:search} lists the search runs on the validation data of \splitR{}.
We started with a feasibility run on 4,000 apps (4k), searched the cross-view SupCon weight, the label type, the dynamic weight and the learning rate on 4k, then the pooling and the static weight on 16,000 apps (16k), and confirmed the final settings on the full set (36k).
The selection score was the mean of the three validation AUCs.
More training data gave the largest gain, especially for the dynamic view (0.954 at 4k, 0.975 at 16k, 0.982 at 36k).
In the search, adding family labels to SupCon lowered the mean validation AUC (0.972 against 0.977 at $\lambda_x=1$).
Mean pooling gave a dynamic validation AUC of 0.975 against 0.972 for last-token pooling on the 16k build.
Fig.~\ref{fig:training} shows the training loss and the validation AUC of the final \detm{} run.

\begin{algorithm}[!htbp]
\caption{Building the two text views of an app}
\label{alg:inputs}
\begin{algorithmic}[1]
\Require APK $a$, DroidBox report $r$, document frequency $\mathrm{df}$ of APIs on the training apps, budgets $B_s{=}2048$, $B_d{=}1024$, view instructions $p_s$, $p_d$ (Table~\ref{tab:hparams} of the paper)
\State $D \gets$ classes defined in the DEX files of $a$
\State $A \gets \{\textsc{Short}(c.m) : (c,m) \in \textsc{MethodRefs}(a),\ c \text{ has a core framework prefix},\ c \notin D\}$
\If{$\textsc{Tokens}(A) > B_s$} \Comment{keep the rarest APIs first}
  \State $A \gets$ the longest prefix of $\textsc{SortBy}(A, \mathrm{df})$ that fits $B_s$
\EndIf
\State $x^{s} \gets p_s \,\Vert\, \textsc{Join}(\textsc{Sort}(A))$ \Comment{one API per line}
\State $E \gets [\,\textsc{Line}(e) : e \in \textsc{Events}(r)\,]$ \Comment{category, operation, target}
\State $E \gets$ \textsc{Normalize}($E$) \Comment{remove hashes, time, ids; \texttt{<pkg>}, \texttt{<hex>}, \texttt{<n>}}
\State $x^{d} \gets p_d \,\Vert\, \textsc{Bag}(E)$ \Comment{unique lines with counts \texttt{x$N$}}
\State \Return $x^{s}$, $x^{d}$ \Comment{the encoder truncates $x^d$ to $B_d$ tokens}
\end{algorithmic}
\end{algorithm}
\begin{algorithm}[!htbp]
\caption{One training step of \name{} (global batch of $B$ apps on $G$ GPUs)}
\label{alg:train}
\begin{algorithmic}[1]
\Require encoder $f_{\theta}$ (frozen base, trainable LoRA $\theta$), batch $\{(x^{s}_i, x^{d}_i, y_i)\}_{i=1}^{B}$, weights $\lambda_x,\lambda_d,\lambda_s$, temperature $\tau$
\State on each GPU $g$: encode its $B/G$ apps: $z^{s}_i \gets f_{\theta}(x^{s}_i)$, $z^{d}_i \gets f_{\theta}(x^{d}_i)$ \Comment{mean pool, L2 norm}
\State gather $Z^{s}=[z^{s}_1..z^{s}_B]$, $Z^{d}=[z^{d}_1..z^{d}_B]$ from all GPUs \Comment{keep the local gradients}
\State $S \gets Z^{s}Z^{d\top}/\tau$
\State $\mathcal{L} \gets \tfrac12\big[\mathrm{CE}(S, \mathrm{diag}) + \mathrm{CE}(S^{\top}, \mathrm{diag})\big]$ \Comment{Eq.~(\ref{eq:nce}) of the paper}
\State $\mathcal{L} \gets \mathcal{L} + \lambda_x\,\mathcal{L}_{\mathrm{SC}}([Z^{s};Z^{d}],[y;y]) + \lambda_d\,\mathcal{L}_{\mathrm{SC}}(Z^{d},y) + \lambda_s\,\mathcal{L}_{\mathrm{SC}}(Z^{s},y)$ \Comment{Eq.~(\ref{eq:supcon}) of the paper}
\State backpropagate $\mathcal{L}$; all-reduce the LoRA gradients; clip; AdamW step on $\theta$
\end{algorithmic}
\end{algorithm}
\begin{algorithm}[!htbp]
\caption{Uses of the fingerprints (no further training of the encoder)}
\label{alg:use}
\begin{algorithmic}[1]
\Require trained encoder $f$; labelled training fingerprints; test app with one or both views
\Statex \textbf{Detection (one view $v$):} train LR on $\{(z^{v}_i,y_i)\}$; score $p(\text{malware}\mid z^{v})$
\Statex \textbf{Fusion:} train LR on $\{([z^{s}_i;z^{d}_i],y_i)\}$; score $[z^{s};z^{d}]$
\Statex \textbf{Transfer \sd{} (\ds{}):} train LR on static (dynamic) fingerprints; score the dynamic (static) fingerprint
\Statex \textbf{Matching \ds{}:} rank the gallery $\{z^{s}_j\}$ by $z^{d}\!\cdot z^{s}_j$; report the rank of the matching APK
\Statex \textbf{Verification:} accept the pair $(r, a)$ if $z^{d}_r\!\cdot z^{s}_a > t$
\Statex \textbf{Family attribution:} similarity-weighted family vote of the $k{=}10$ nearest $z^{s}_j$ of a labelled malware gallery
\end{algorithmic}
\end{algorithm}
\begin{table}[!htbp]
\caption{Hyperparameter search (validation data of \splitR{}; each row is one run). $\lambda_x,\lambda_d,\lambda_s$: SupCon weights of Eq.~(\ref{eq:total}) of the paper; ``bin+fam'': SupCon with binary and family labels. Pool: L = last token, M = mean. The validation sets differ in size between the builds (600, 2,400 and 5,441 apps).}
\label{tab:search}
\centering
\small
\begin{tabular}{@{}l c c c c c c c@{}}
\toprule
Build & $\lambda_x$ & $\lambda_d$ & $\lambda_s$ & lr & Pool & Ep. & Static / dynamic / fusion AUC \\
\midrule
4k (\frz{}) & -- & -- & -- & -- & L & -- & 0.952 / 0.947 / 0.971 \\
4k (feasibility) & 0.5 & 0 & 0 & $10^{-4}$ & L & 2 & 0.979 / 0.938 / 0.988 \\
4k & 1 & 0 & 0 & $10^{-4}$ & L & 3 & 0.989 / 0.947 / 0.994 \\
4k & 2 & 0 & 0 & $10^{-4}$ & L & 3 & 0.987 / 0.952 / 0.992 \\
4k, bin+fam & 1 & 0 & 0 & $10^{-4}$ & L & 3 & 0.984 / 0.940 / 0.991 \\
4k & 2 & 2 & 0 & $10^{-4}$ & L & 3 & 0.986 / 0.954 / 0.993 \\
4k & 2 & 0 & 0 & $2\times10^{-4}$ & L & 3 & 0.992 / 0.954 / 0.997 \\
16k & 2 & 2 & 0 & $2\times10^{-4}$ & L & 2 & 0.994 / 0.972 / 0.996 \\
16k & 2 & 2 & 0 & $2\times10^{-4}$ & M & 2 & 0.994 / 0.975 / 0.997 \\
16k & 2 & 2 & 2 & $2\times10^{-4}$ & L & 2 & 0.995 / 0.970 / 0.996 \\
36k & 2 & 2 & 2 & $2\times10^{-4}$ & M & 2 & 0.995 / 0.981 / 0.998 \\
\textbf{36k (final, \detm{})} & 2 & 2 & 4 & $2\times10^{-4}$ & M & 3 & \textbf{0.996 / 0.982 / 0.998} \\
36k & 2 & 2 & 8 & $2\times10^{-4}$ & M & 4 & 0.995 / 0.982 / 0.998 \\
\bottomrule
\end{tabular}
\end{table}

\clearpage
\section{Additional results}
\label{app:extra}

\paragraph*{Training variance}
Table~\ref{tab:seeds} of the paper gives the three seeds of the final configuration on \splitR{}.
Seed 0 is the run that the search selected.
The detection and transfer AUCs vary by at most 0.004 between the seeds.
The matching varies more: \Rk{10} goes from 0.089 to 0.137, and seed 0 has the highest value.
All seeds stay far below \matm{} (0.466).
Table~\ref{tab:f1} gives the secondary F1 metric for seed 0.

\begin{figure}[!htbp]
\centering
\includegraphics[width=0.85\linewidth]{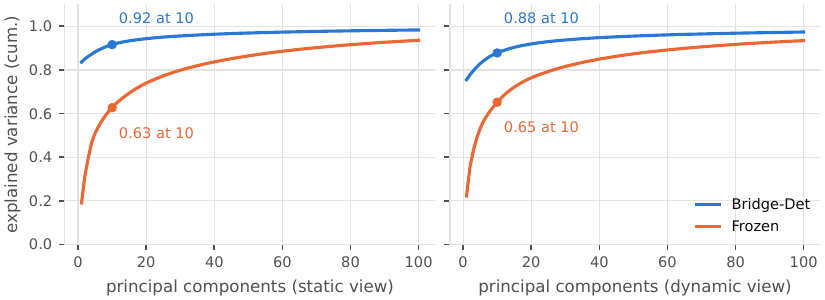}
\caption{Cumulative explained variance of the fingerprints (\splitR{} test, \detm{} seed 0). 10 components explain 92\% (static) and 88\% (dynamic) of the variance, against 63\% and 65\% for \frz{}.}
\label{fig:pca}
\end{figure}

\begin{table}[!htbp]
\caption{Secondary metric: F1 at threshold 0.5 (\splitR{} test, LR probe, \detm{} seed 0). F1 was read after the test.}
\label{tab:f1}
\centering
\small
\begin{tabular}{@{}l c c c@{}}
\toprule
Model & Static & Dynamic & Fusion \\
\midrule
\detm{} & 0.980 & 0.932 & 0.982 \\
\frz{} & 0.938 & 0.906 & 0.957 \\
\bottomrule
\end{tabular}
\end{table}
\paragraph*{RQ1}
Fig.~\ref{fig:pca} shows that the fine-tuned space is concentrated in a few directions, which matches the loss of app identity under SupCon.
Fig.~\ref{fig:matching_family} of the paper gives the matching curves on \splitF{}.

\begin{figure}[!htbp]
\centering
\includegraphics[width=0.85\linewidth]{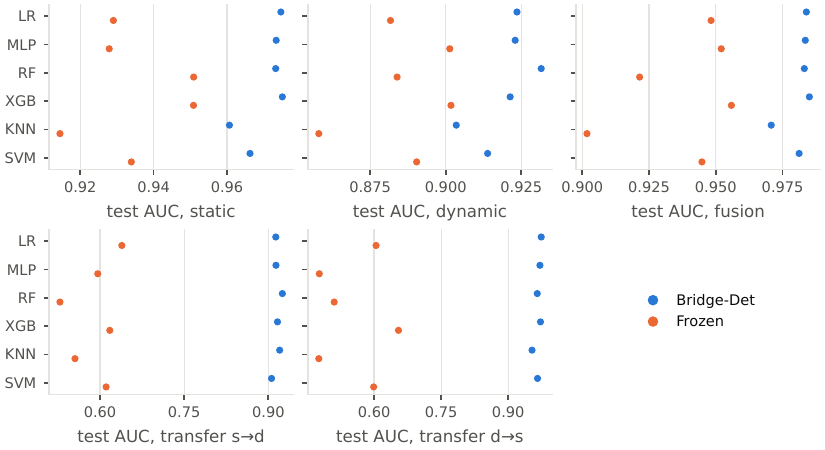}
\caption{Test AUC of six classifiers on five inputs on \splitF{}. Each classifier and its hyperparameters are chosen on the validation data.}
\label{fig:classifiers_family}
\end{figure}

\paragraph*{RQ2}
Fig.~\ref{fig:classifiers_family} repeats the six-classifier comparison of Fig.~\ref{fig:classifiers} of the paper on \splitF{}.

\paragraph*{RQ3}
Table~\ref{tab:degrade} gives the values of the controlled degradation of Fig.~\ref{fig:degradation} of the paper.

\begin{figure}[!htbp]
\centering
\includegraphics[width=0.85\linewidth]{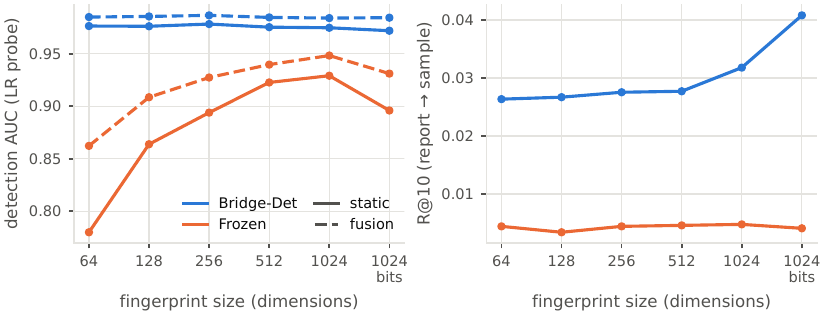}
\caption{Compact fingerprints on \splitF{}: detection AUC and matching \Rk{10} against the fingerprint size, from 64 float dimensions to 1,024 sign bits.}
\label{fig:compact_family}
\end{figure}

\begin{table}[!htbp]
\caption{Controlled, random degradation (\splitR{} test, AUC, \detm{} seed 0). ``View'': the degraded view alone. ``Fusion'': the degraded view together with the clean other view.}
\label{tab:degrade}
\centering
\small
\begin{tabular}{@{}l c c c c@{}}
\toprule
 & \multicolumn{2}{c}{\detm{}} & \multicolumn{2}{c}{\frz{}} \\
\cmidrule(lr){2-3}\cmidrule(l){4-5}
Test change & View & Fusion & View & Fusion \\
\midrule
None (static / dynamic view)    & 0.995 / 0.978 & 0.996 & 0.978 / 0.962 & 0.989 \\
50\% of API lines removed       & 0.985 & 0.992 & 0.932 & 0.976 \\
90\% of API lines removed       & 0.879 & \textbf{0.973} & 0.772 & 0.901 \\
5 dynamic events kept           & 0.891 & 0.994 & 0.827 & 0.959 \\
2 dynamic events kept           & 0.680 & \textbf{0.992} & 0.640 & 0.917 \\
\bottomrule
\end{tabular}
\end{table}
\paragraph*{RQ5}
Fig.~\ref{fig:compact_family} repeats the compact fingerprints of Fig.~\ref{fig:compact} of the paper on \splitF{}; Fig.~\ref{fig:labels} of the paper gives the label efficiency.
With $n=10$ labelled apps per class, the static AUC on \detm{} fingerprints is 0.995, while \frz{} gives 0.72.
This result shows the quality of the space, not a low-label method: the encoder itself was trained with all training labels.

\begin{figure}[!htbp]
\centering
\includegraphics[width=0.85\linewidth]{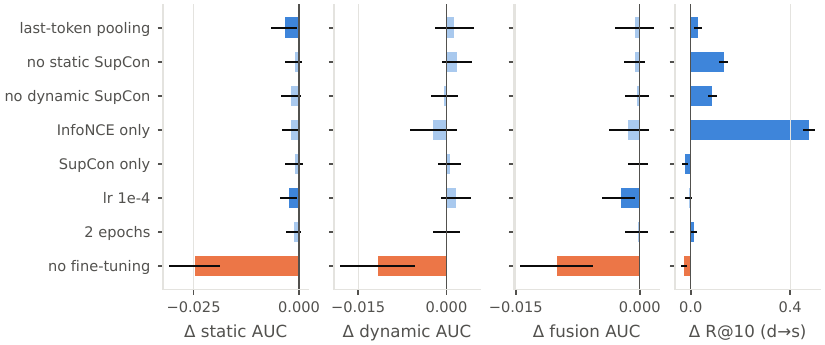}
\caption{Ablation effects (run $-$ final configuration) with 95\% CIs on the 16k subset. A light bar means that the CI includes zero.}
\label{fig:ablation}
\end{figure}

\begin{figure}[!htbp]
\centering
\includegraphics[width=0.65\linewidth]{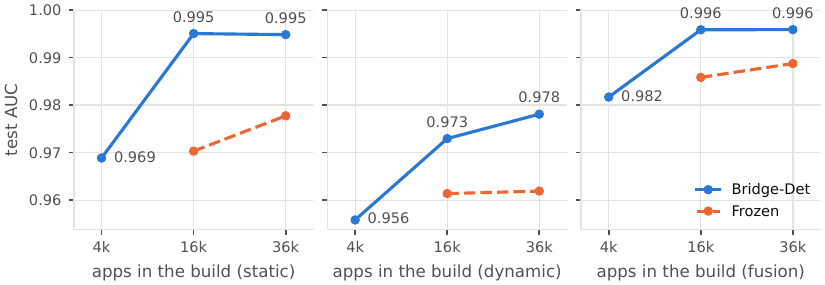}
\caption{Test AUC of the final configuration against the size of the training build (each build tested on its own \splitR{} test part).}
\label{fig:datasize}
\end{figure}

\paragraph*{Ablation and data size}
Fig.~\ref{fig:ablation} plots the effects of Table~\ref{tab:ablation} of the paper with their 95\% CIs.
Fig.~\ref{fig:datasize} gives the test AUC of the final configuration against the size of the training build.
The test sets differ between the builds, so these values are not paired; on the 84 test apps that the 4k and 16k builds share, the 16k model is significantly better for static detection and fusion, and the dynamic difference is not significant.


\section{Reproducibility}
\label{app:repro}

All tools and Python packages come from a pinned Nix environment (nixpkgs commit \texttt{ac62194c}, Python 3.12, PyTorch 2.7.0 with CUDA 12.8, Transformers 4.51.3, PEFT 0.15.2, scikit-learn 1.6.1, NumPy 2.2.5).
We trained on 8 NVIDIA Quadro RTX 4000 GPUs (8\,GB each) with a micro-batch of 4 apps per GPU, fp16 mixed precision and gradient checkpointing; one run of 3 epochs (2,379 steps) took 3.2--3.4 hours.
The preprocessing reads the APKs and the DroidBox reports from the data archives in memory; it does not run any app.
All 18,140 benign apps and 18,078 of the 18,136 malware apps are in AndroZoo; the released split file gives the SHA-256 hash of every app and a script that downloads the APKs from AndroZoo by hash. The other 58 malware apps are available from the MalDozer authors.
The training host stopped several times because of GPU communication time-outs; every run resumed from its last checkpoint (saved every 20 steps), so the data order after a resume is not identical to that of a run without a stop.
The repository \url{https://github.com/Mouatez/bridge-static-dynamic} holds the code, the configurations, the result files and the dataset of the paper: \texttt{data/metadata.csv} gives the SHA-256, SHA-1 and MD5 hashes, the AndroZoo metadata, the labels, the families and the splits of all 36,276 apps, so that the APKs can be downloaded from AndroZoo; its release v1.0 holds the derived text inputs, the trained adapters and the test fingerprints, with SHA-256 checksums. With the repository and the release, all results of the paper are fully reproducible.

\end{document}